\documentclass[onecolumn]{revtex4}
\usepackage{amssymb}

\def\al{\alpha}
\def\be{\beta}
\def\ga{\gamma}
\def\de{\delta}
\def\ep{\epsilon}

\def\et{\eta}

\def\ka{\kappa}
\def\la{\lambda}

\def\si{\sigma}

\def\ph{\phi}

\def\ps{\psi}
\def\om{\omega}
\def\Ga{\Gamma}

\def\La{\Lambda}

\def\fr#1#2{{{#1} \over {#2}}}
\def\half{{\textstyle{1\over 2}}}

\def\frac#1#2{{\textstyle{{#1}\over {#2}}}}

\def\vev#1{\langle {#1}\rangle}

\def\lsim{\mathrel{\rlap{\lower4pt\hbox{\hskip1pt$\sim$}}
    \raise1pt\hbox{$<$}}}
\def\gsim{\mathrel{\rlap{\lower4pt\hbox{\hskip1pt$\sim$}}
    \raise1pt\hbox{$>$}}}
\def\sqr#1#2{{\vcenter{\vbox{\hrule height.#2pt
         \hbox{\vrule width.#2pt height#1pt \kern#1pt
         \vrule width.#2pt}
         \hrule height.#2pt}}}}

\def\pt#1{\phantom{#1}}

\def\nsc#1#2#3{\om_{#1}^{{\pt{#1}}#2#3}}
\def\lsc#1#2#3{\om_{#1#2#3}}
\def\usc#1#2#3{\om^{#1#2#3}}
\def\lulsc#1#2#3{\om_{#1\pt{#2}#3}^{{\pt{#1}}#2}}

\def\vb#1#2{e_{#1}^{{\pt{#1}}#2}}
\def\ivb#1#2{e^{#1}_{{\pt{#1}}#2}}
\def\uvb#1#2{e^{#1#2}}
\def\lvb#1#2{e_{#1#2}}

\newcommand{\beq}{\begin{equation}}
\newcommand{\eeq}{\end{equation}}
\newcommand{\bea}{\begin{eqnarray}}
\newcommand{\eea}{\end{eqnarray}}
\newcommand{\bit}{\begin{itemize}}
\newcommand{\eit}{\end{itemize}}
\newcommand{\rf}[1]{(\ref{#1})}

\begin{document}

\title{Spontaneous and Explicit Spacetime Symmetry Breaking \\
in Einstein-Cartan Theory with Background Fields}

\author{Robert Bluhm and Yu Zhi}

\affiliation{
Physics Department, Colby College,
Waterville, ME 04901  USA
}

\begin{abstract}
Explicit and spontaneous breaking of spacetime symmetry
under diffeomorphisms, local translations, and local Lorentz transformations
due to the presence of fixed background fields is examined in Einstein-Cartan theory.
In particular, the roles of torsion and violation 
of local translation invariance are highlighted.
The nature of the types of background fields that can arise and how they cause
spacetime symmetry breaking is discussed.
With explicit breaking,
potential no-go results are known to exist,
which if not evaded lead to inconsistencies between the Bianchi identities,
Noether identities, and the equations of motion.
These are examined in detail,
and the effects of nondynamical backgrounds and explicit breaking
on the energy-momentum tensor when torsion is present are discussed as well.
Examples illustrating various features of both explicit and spontaneous breaking
of local translations are presented and compared to the case of diffeomorphism breaking.
\end{abstract}

\maketitle

\section{Introduction}

Local spacetime symmetries are fundamental features of current theories of gravity,
including Einstein's General Relativity (GR).
In GR the underlying geometry is Riemann,
which is characterized by the Riemann curvature tensor.
In the Einstein equations, the energy-momentum density of the matter fields 
acts as the source of the curvature.

An extension of GR, which is useful in describing fields with spin,
is Einstein-Cartan (EC) 
theory~\cite{cartan1,cartan2,sciama62,TWBK61,HehlRMP76,HehlPR95,Blag02,at06,bg13,PBO17,yo23}.
In this case, the underlying geometry is Riemann-Cartan,
which is characterized by both a curvature and a torsion tensor.
Using a vierbein formalism, 
the independent fields are the vierbein and the spin connection.
The pure-gravity term in the action has an Einstein-Hilbert form,
but in this case a spin density due to the presence of spin fields
acts as the source of torsion while the Einstein equations 
couple energy-momentum and curvature.
The torsion in EC theory is fixed by the spin density and is zero in
regions of spacetime where the spin density vanishes.
Since torsion couples to spin density only very weakly,
no experiments have detected it~\cite{ilspr02,aknrjt08}.
Despite the lack of evidence for torsion, 
EC theory can be viewed as a viable alternative to GR 
that has the advantage of incorporating spin in a straightforward manner.

In one approach to EC theory, it is common to consider invariance under 
diffeomorphisms (Diffs) in a spacetime frame as well as
invariance under local Lorentz transformations (LLTs) in a local 
Lorentz basis as fundamental spacetime symmetries.
In GR, the spin connection is completely determined by the vierbein,
while in EC theory, the vierbein and spin connection are independent
of each other when the torsion is nonzero.
However, the spin connection does not propagate as 
independent degrees of freedom in EC theory.
There are also generalizations that go beyond EC theory,
which contain additional terms in the pure-gravity action 
that allow the spin connection to propagate.
However, in such extensions,
questions concerning unphysical ghost modes, negative energies,
or discrepancies with observations
must be resolved, and the differences with GR are greater.

In many respects, local spacetime symmetries are similar to local gauge symmetries,
where the latter are fundamental in the Standard Model (SM) in particle physics.
This has led to considerable interest in EC theory formulated as a gauge theory,
where Poincare symmetry is treated as a local gauge
symmetry~\cite{sciama62,TWBK61,HehlRMP76,HehlPR95,Blag02,at06,bg13,PBO17,yo23}.
In this context, the fundamental spacetime symmetries consist of LLTs
and local translations (LTs),
and the vierbein becomes the gauge field for the LTs 
while the spin connection is the gauge field for the LLTs.
At the same time, a theory of gravity must be covariant, 
which implies invariance under Diffs, 
as long as the theory contains only physical dynamical fields.
In the end, it is largely a matter of choice whether to consider 
Diffs and LLTs as fundamental
versus considering LTs and LLTs in this way.

In parallel with constructing EC theory and its generalizations
as gauge theories with local Poincare invariance,
much effort has also been devoted 
to understanding how gravity can be quantized
and how the effects of this might be discovered.
One idea that has been investigated widely is that in a quantum theory
of gravity small violations of spacetime symmetry can occur.
Mechanisms for how this might occur can be found, for example,
in string theory~\cite{akss1,akss2,akss3}.
A phenomenological framework known as the Standard-Model Extension (SME)
has been developed, which is useful in exploring the possibility of spacetime 
symmetry breaking~\cite{sme1,sme2,sme3,akgrav04,rbsme,JT14,RB14,hees2016}.
It is based on the general idea idea that no matter how such 
breakings might occur,
the effects of these violations should be describable in the context of 
an effective field theory that contains both the SM and EC theory at low energies.

The SME is constructed by adding to the action any possible interaction terms that 
involve couplings with SM or gravitational fields that break spacetime 
symmetry while maintaining observer independence.
Such interactions introduce fixed background fields, usually referred to
as SME coefficients, which couple with the SM and gravitational fields.
Using the SME framework, many experimental searches for spacetime
symmetry breaking have been conducted with extremely high sensitivities
over the past several decades~\cite{aknr-tables}.  

When the gravitational sector of the SME was first investigated it was
found that there are fundamental differences depending on whether spacetime
symmetry breaking occurs spontaneously or explicitly~\cite{akgrav04}.
Due to the nondynamical nature of the backgrounds with explicit breaking,
inconsistencies between the Bianchi identities and equations of motion can occur,
which in a Riemann geometry create conflicts with covariant energy-momentum conservation.
However, these results, known as no-go results, do not occur with spontaneous breaking.
For this reason it was typically assumed when using the gravity sector of the SME
that spacetime symmetry breaking occurs spontaneously~\cite{qbak06,akjt1,akjt2}.
Specifically, it was spontaneous breaking of Diffs and LLTs that 
was most widely investigated,
and questions concerning Nambu-Goldstone (NG) bosons, massive Higgs-like fields, 
and the possibility of a gravitational Higgs mechanism were 
examined~\cite{bb1,bb2}.
In addition, linkage between spontaneous breaking of Diffs and LLTs was found,
in that when vacuum values exist that spontaneously break Diffs, 
vacuum values also exist that spontaneously break LLTs, and vice versa.
Vector models known as Bumblebee models were studied as examples that
illustrate these and other results of spontaneous breaking of Diffs and 
LLTs~\cite{akss3,akgrav04,bb1,bb2,rbngrpav}.

In some subsequent works,
explicit breaking of Diffs and LLTs,
due to the presence of fixed backgrounds,
was examined in more detail,
and it was found that in some cases the no-go results can be 
evaded~\cite{rb15a,rb15b,rbas16,rbSym17}.
Noether identities that must hold as a results of observer independence were shown to 
provide a useful tool for determining if a particular model must be ruled out or not.
It was also found that in some cases a hybrid form of spacetime symmetry breaking
can occur,  involving both explicit and spontaneous breaking.
The question of whether the SME can accommodate explicit spacetime
symmetry breaking was reexamined,
and it was shown that in some cases the potential
no-go results can be evaded and the SME can be 
used with explicit breaking~\cite{rbhbyw19,rbyySym21}.
For simplicity, effects of torsion were largely ignored in much of these works,
and LTs were not considered.

However, in~\cite{ybcc18a,ybcc18b,ccyb19,ybcp20,bhr23},
the roles of torsion and LTs  were considered in more detail in gravity 
theories with explicit-breaking SME coefficients, 
and some interesting properties were found.
For example, it was shown that when a nondynamical background field is 
present in EC theory,
the torsion can be nonzero in regions where there is no spin density
associated with matter.
Linkages between explicit breaking of LTs with explicit breaking 
of Diffs and LLTs were explored as well.

All of this led to a complete generalization of the SME being developed,
which includes nondynamical backgrounds that can
explicitly break Diffs and LLTs~\cite{akzl21a}.
However, for simplicity, many of the effects of torsion were again largely ignored
and LTs were not directly considered. 
Nonetheless, possible linkages between explicit breaking of Diffs and LLTs,
spontaneous breaking of these symmetries, or hybrid combinations of both
types of breakings were catalogued and investigated.
Applications of how this new approach can be used and some examples of 
tests and their sensitivities are described in~\cite{akzl21b}.

Ultimately, any theory with explicit breaking that does 
not evade the no-go results must be ruled out in Riemann geometry 
or in Riemann-Cartan geometry if torsion is included.
An idea that has been widely explored is that these theories
might instead be consistent in a Finsler geometry or some other beyond-Riemann 
geometry~\cite{BCS2000,akplb10,akplb11,aknrrt12,dcpm12,jsca14,nr15,ms15a,ms15b,jfrl15,clvp18}.
Based on this,
the interpretation when working with the generalization of the SME that includes explicit breaking is 
that any detection of spacetime symmetry breaking 
involving interactions that do not evade the no-go results
would indicate the existence of a beyond-Riemann geometry, 
such as Finsler geometry~\cite{akzl21a,akzl21b}.

The primary goals of this paper are to revisit EC theory when background fields 
that explicitly or spontaneously break spacetime symmetries are present and to elaborate on
and fill in certain features or possibilities that have largely or partially been ignored.
A general formalism containing a variety of different types of backgrounds is used.
In particular, breaking of all three spacetime symmetries,
Diffs, LTs, and LLTs, and the linkages between them, 
are examined for both explicit and spontaneous breaking with torsion included.
In each case, the question of whether no-go results can appear is addressed,
and implications concerning covariant conservation of the energy-momentum tensor 
with torsion present are examined.
Specific examples of how LTs are broken either explicitly or spontaneously when torsion
is present are provided.

The organization of this paper is as follows:
Section~\ref{sec2} gives background on EC Thoery
for the usual case of when Diffs, LTs, and LLTs are not broken.
Readers already familiar with EC theory may want to
skip ahead and start with Section~\ref{sec3},
which then looks at how these symmetries,
are broken either spontaneously or explicitly
when background fields are present.
Section~\ref{sec4} looks at explicit breaking in detail,
including the no-go results and Noether identities that follow
from observer independence.
Spontaneous breaking is examined in Section~\ref{sec5},
and examples of bumblebee theories with spontaneous breaking of
LTs with torsion included are presented.
Section~\ref{sec6} provides some discussion and conclusions.
The notation and conventions used here follow those in~\cite{akgrav04}.

%%%%%%%%%%%%%%%%%%%%%%%%%%%%%%%%%%%%%%%%%%%%%%%%%%%%%%

\section{EC Theory} \label{sec2}

In EC theory,
there is both curvature and torsion,
and it is assumed that the nonmetricity vanishes
so that $D_\la g_{\mu\nu} = 0$.
A vierbein formalism can be used,
where $\vb \mu a$ is the vierbein and
$\nsc \mu ab$ is the spin connection.
In this notation, Greek letters denote components 
with respect to the spacetime frame,
while Latin letters denote components with 
respect to the local Lorentz basis.
The metric is given in terms of the vierbein 
as $g_{\mu\nu} = \vb \mu a \vb \nu b \et_{ab}$,
where $\et_{ab}$ is the local Minkowski metric.
The connection is the Cartan connection $\Ga^\la_{\pt{\la}\mu\nu}$,
which has an antisymmetric part that defines the torsion tensor:
\beq
T^\la_{\pt{\la}\mu\nu} = \Ga^\la_{\pt{\la}\mu\nu} - \Ga^\la_{\pt{\la}\nu\mu} \, .
\label{Torsiondef}
\eeq
Using a tilde to denote the symmetric Levi-Civita connection in GR,
which has components given by the Christoffel symbol,
\beq
\tilde \Ga^\la_{\pt{\la}\nu\mu} = \left\{ \begin{array}{c} \la \\ \mu \nu \end{array} \right\} \, ,
\label{tildeTChrist}
\eeq
and defining the contorsion tensor as
\beq
K^{\la\mu\nu} = \fr 1 2 (T^{\la\mu\nu} - T^{\mu\nu\la} - T^{\nu\mu\la}) \, ,
\label{contorsionK}
\eeq
the Cartan connection can be written as
\beq
\Ga^\la_{\pt{\la}\mu\nu} = \tilde \Ga^\la_{\pt{\la}\nu\mu} + K^\la_{\pt{\la}\nu\mu} \, .
\label{Torsiondef2}
\eeq
Assuming the covariant derivative of the vierbein vanishes gives a relation
involving the spin connection $\nsc \mu ab$ as
\beq
D_\mu \vb \nu a = \partial_\mu \vb \nu a - \Ga^\la_{\pt{\la}\mu\nu} \vb \la a + \lulsc \mu a b \vb \nu b = 0 \, .
\label{De=0}
\eeq
With this, the connection and torsion can be found in
terms of the vierbein and spin connection:
\beq
\Ga^\la_{\pt{\la}\mu\nu} = \uvb \la a (\partial_\mu \lvb \nu a - \lulsc \mu b a \lvb \nu b ) \, ,
\label{Gammavierb}
\eeq
\beq
T_{\la\mu\nu} = \vb \la a [ (\partial_\mu \lvb \nu a + \lsc \mu a b \, \vb \nu b) 
- (\mu \leftrightarrow \nu ) ] \, .
\label{Torvieb}
\eeq

The curvature tensor is defined as
\beq
R^\ka_{\pt{\ka}\la\mu\nu} = \partial_\mu \Ga^\ka_{\pt{\ka}\nu\la} - \Ga^\ka_{\pt{\ka}\mu\si} \Ga^\si_{\pt{\si}\nu\la}
- (\mu \leftrightarrow \nu ) \, ,
\label{curvR}
\eeq
and its contractions, 
\beq
R_{\mu\nu} = R^\ka_{\pt{\ka}\mu\ka\nu} \, , \quad R = g^{\mu\nu} R_{\mu\nu} \, ,
\label{RicciR}
\eeq
give, respectively, the Ricci tensor and the curvature scalar.
Note that in EC theory, 
the Ricci tensor is not symmetric and neither is the Einstein tensor,
$G^{\mu\nu} = R^{\mu\nu} - \half g^{\mu\nu} R$.
The curvature can also be given in terms of the vierbein and spin connection as
\beq
R^\ka_{\pt{\ka}\la\mu\nu}= \ivb \ka a \vb \la b [ (\partial_\mu \lulsc \nu a b + \lulsc \mu a c \lulsc \nu c b) 
- (\mu \leftrightarrow \nu ) ] \, .
\label{Rvieb}
\eeq

Bianchi identities for the curvature and torsion in EC theory are off-shell geometric identities,
which are given as:
\bea
\sum_{(\la \mu \nu)} [ D_\nu R^\al_{\pt{\al}\be\la\mu} + T^\si_{\pt{\si}\la\mu} R^\al_{\pt{\al}\be\si\nu} ]= 0 \, , 
\label{Bianchi1} \\
\sum_{(\la \mu \nu)} [ D_\nu T^\al_{\pt{\al}\la\mu} + T^\si_{\pt{\si}\la\mu} T^\al_{\pt{\al}\si\nu} - R^\al_{\pt{\al}\nu\la\mu}]  = 0 \, , 
\label{Bianchi2} 
\eea
where the sum in each case is over the cyclic permutations of $(\la \mu \nu)$.
Two useful off-shell identities can be derived from these by contracting and manipulating terms~\cite{akgrav04}.
The results are
\bea
(D_\mu - T^\la_{\pt{\la}\la\mu}) G^{\mu\nu} + T_{\la\mu}^{\pt{\la\mu}\nu} G^{\mu\la} 
- \half R^{\al\be\mu\nu} \hat T_{\mu\al\be} = 0  \, ,
\label{ContractedBianchi1}
\\
G^{\mu\nu} - G^{\nu\mu} = - (D_\si - T^\la_{\pt{\la}\la\si}) \hat T^{\si\mu\nu}  \, .
\quad\quad\quad
\label{ContractedBianchi2}
\eea

\subsection{EC Action}

The generic form of the action can be written in terms of the vierbein, spin connection, and matter fields as
\beq
S = S_g + S_{g,m} 
= \fr 1 {2 \ka} \int d^4x e \, R (\vb \mu a,\nsc \mu a b) +  \int d^4x e {\cal L}_m (\vb \mu a,\nsc \mu a b, f^\ps) \, .
\label{ECS}
\eeq
Here, $S_g$ is the Einstein-Hilbert term, with the curvature scalar expressed as a function
of the vierbein and the spin connection.
The matter term $S_{g,m}$ depends on the vierbein, spin connection, and matter fields,
where the latter are denoted generically as $f^\ps$.
The specific forms and suitable indices for $f^\ps$ depend on the types of fields that are included,
which can consist of tensor and spin fields.
The coupling $\ka = 8 \pi G$ (with $c=1$), 
and $e$ is the determinant of the vierbein.

Note that a cosmological constant term could also be added to the action,
and as an alternative to EC theory, kinetic terms for the torsion would be added as well.
This would permit inclusion of teleparallel gravity when the curvature vanishes~\cite{mh23,jhk2019}.
However, these considerations go beyond the scope of this work,
which for simplicity considers only flat spacetime background in vacuum.  

Variation of the action term $S_{g,m} $ with respect to $\vb \mu a $, $\nsc \mu a b$, and $f^\ps$ 
has the form:
\beq
\de S_{g,m}  = \int d^4x e \, \left[ T_e^{\mu\nu} e_{\nu a} \, \de \vb \mu a 
+  \half S_{\om \, \pt{\mu} a b}^{\pt{\om \, }\mu} \, \de \nsc \mu a b + \fr {\de S_{g,m} } {\de f^\ps} \de f^\ps \right] \, ,
\label{ECdeltaSm}
\eeq
which defines $T_e^{\mu\nu}$ as the energy-momentum tensor and $S_{\om \, \pt{\mu} a b}^{\pt{\om \, }\mu}$
as the spin density.
Varying the matter fields gives the Euler-Lagrange expression for $f^\ps$.

When the full action is varied with respect to the dynamical fields, the result is:
\beq
\de S = \int d^4x e \, \left[ (-\fr 1 {\ka} G^{\mu\nu} + T_e^{\mu\nu}) e_{\nu a} \, \de \vb \mu a 
+ (\fr 1 {2 \ka} \hat T^{\la\mu\nu} e_{\mu a} e_{\nu b} + \half S_{\om \, \pt{\mu} a b}^{\pt{\om \, }\mu}) \, \de \nsc \mu a b
+ \fr {\de S_{g,m} } {\de f^\ps} \de f^\ps \right] \, ,
\label{ECdeltaS}
\eeq
where 
\beq
\hat T^{\la\mu\nu} = T^{\la\mu\nu}  + T^{\al \pt{\al} \mu}_{\pt{\al} \al} g^{\la\nu} - T^{\al \pt{\al} \nu}_{\pt{\al} \al} g^{\la\mu}
\label{That}
\eeq
is the trace-corrected torsion tensor.
Setting $\de S = 0$ gives the equations of motion for the vierbein, spin connection,
and matter fields, respectively, as
\bea
G^{\mu\nu} = \ka T_e^{\mu\nu} \, ,  \quad\quad
\label{EinsteinEq} \\
\hat T^{\la\mu\nu}  =   - \ka S_{\om \, \pt{\mu} }^{\pt{\om \, } \la\mu\nu} \, ,
\label{TorsionEq} \\
\fr {\de S_{g,m} } {\de f^\ps} = 0 \, .
\quad\quad\quad\,\,\,\,
\label{feq}
\eea

\subsection{Spacetime Symmetries in EC Theory}  \label{sec2c}

The local spacetime symmetries, diffeomorphisms (Diffs), 
local Lorentz transformation (LLTs), and local translations (LTs),
act on $\vb \mu a$, $\nsc \mu a b$, and $f^\ps$
leaving the action invariant.
Under Diffs,
the transformation rules are given as Lie derivatives, 
which can be rewritten using the full covariant derivative $D_\mu$
that corrects with both the Cartan connection $\Ga^\la_{\pt{\la}\mu\nu}$
and the spin connection,
depending on what types of fields it acts on.   
However,
the Poincare algebra that includes LLTs and LTs
uses a Lorentz covariant derivative, $D^{(\om)}_a = \ivb \mu a D^{(\om)}_\mu$,
where $D^{(\om)}_\mu$ corrects with only the spin connection~\cite{HehlRMP76}.
For example,
\beq 
D^{(\om)}_\mu  \vb \nu a = \partial_\mu \vb \nu a + \lulsc \mu a b \vb \nu b \, ,
\label{omDe=0}
\eeq
which does not vanish,
since the term with the Cartan connection in \rf{De=0} is not included.
Notice that when acting on a tensor with only local indices, for example, $A^b$,
it can also be written as
\beq 
D^{(\om)}_\mu  A^b = \vb \nu b D_\mu  A^\nu  \, ,
\label{omDAb=0}
\eeq
where $A^b = \vb \nu b A^\nu$.

The transformation rules for the the vierbein and spin connection are then defined as follows:
Under Diffs with parameters $\xi^\mu$,
\bea
\de_{\rm Diff} \, \vb \mu a &=& {\cal L}_\xi \vb \mu a 
= (\partial_\mu \xi^\al) \vb \al a + \xi^\al \partial_\al \vb \mu a 
\nonumber \\
&=& (D_\mu \xi^\al ) \vb \al a - T^\al_{\pt{\al}\mu\be} \xi^\be \vb \al a - \xi^\al \nsc \al a b e_{\mu b} \, ,
\label{Diffvb}
\\
\de_{\rm Diff} \, \nsc \mu a b &=& {\cal L}_\xi \nsc \mu a b = (\partial_\mu \xi^\al) \nsc \al a b + \xi^\al \partial_a \nsc \mu a b \, ,
\label{Diffsc}
\eea
where $ {\cal L}_\xi $ denotes a Lie derivative along $\xi^\mu$.
Under LTs with parameters $\ep^a$:
\bea
\de_{\rm LT} \, \vb \mu a &=& D_\mu ^{(\om)} \ep^{a} + \ep^b T^a_{\pt{a} b \mu} \, ,
\label{LTvb}
\\
\de_{\rm LT} \, \nsc \mu a b  &=& \ep^c R^{ab}_{\pt{ab}c \mu} \, .
\label
{LTsc}
\eea
Under LLTs with parameters $\ep_a^{\pt{a}b} = - \ep^b_{\pt{b}a}$:
\bea
\de_{\rm LLT} \, \vb \mu a &=& - \ep^a_{\pt{a}b} \vb \mu b \, ,
\label{LLTvb}
\\
\de_{\rm LLT} \, \nsc \mu a b &=& D_\mu ^{(\om)} \ep^{ab}
\nonumber \\
&=& \partial_\mu \ep^{ab} + \lulsc \mu a c \, \ep^{cb} + \lulsc \mu b c \, \ep^{ac} \, .
\label{LLTsc}
\eea

Note that these transformations are not independent, since it can be shown that
with $\xi^\mu = \ep^\mu = \ivb \mu a \ep^a$ and $\ep^{ab} = \ep^\mu \nsc \mu a b$
that they are related by~\cite{HehlRMP76,Blag02,ybcc18a}:
\beq
\de_{\rm Diff} (\xi^\mu) = \de_{\rm LT} (\ep^a) + \de_{\rm LLT} (\ep^{ab}) \, .
\label{transfrel}
\eeq
As a result, symmetry under Diffs and LLTs implies
symmetry under LTs and LLTs, and vice versa.

Dynamical tensors in EC theory can have
components given with respect to either the spacetime frame
or a local basis frame.
For example, a vector field can have spacetime components $A_\mu$
or local components $A_b$,
where these are related by the vierbein:  $A_\mu = \vb \mu b A_b$.
Under Diffs, LLTs, and LTs the 
local frame components, $A_b$, transform, respectively, as
\bea
\de_{\rm Diff} \, A_b = \xi^\nu \partial_\nu A_b \, , \quad\quad
\nonumber \\
\de_{\rm LLT} \, A_b = - \ep_b^{\pt{b} c} A_c \, , \quad\quad
\nonumber \\
\de_{\rm LT} \, A_b = \ep^c \ivb \mu c D^{(\om)}_{\, \mu} A_b \, .
\label{Ab}
\eea
Making the substitutions $\xi^\mu = \ep^\mu = \ivb \mu c \ep^c$ 
and $\ep^{bc} = \ep^\mu \nsc \mu b c$,
it follows that Eq.\ \rf{transfrel} holds for these transformations.
These symmetry transformations can also be performed on the 
components $A_\mu$ defined with respect to the spacetime frame.
The fact that $A_b$ and the vierbein transform 
results in transformations of $A_\mu$ as well,
which are given under Diffs, LLTs, and LTs, respectively, as
\bea
\de_{\rm Diff} \, A_\mu = (\partial_\mu \xi^\nu) A_\nu + \xi^\nu \partial_\nu A_\mu \, , 
\quad\quad\quad\quad\quad\quad\quad\quad\quad
\nonumber \\
\de_{\rm LLT} \, A_\mu = 0 \, , 
\quad\quad\quad\quad\quad\quad\quad\quad\quad\quad\quad\quad\quad\quad\quad\quad\quad
\nonumber \\
\de_{\rm LT} \, A_\mu = (D^{(\om)}_{\, \mu} \ep^b) \ivb \nu b A_\nu
+ \ep^b  \ivb \nu b  D_{\, \nu} A_\mu + \ep^b T^\nu_{\pt{a} b \mu} A_\nu \, .
\label{Amu}
\eea
Notice that in this case with $\xi^\nu = \ep^\nu = \ivb \nu c \ep^c$, 
it follows that $\de_{\rm Diff} \, A_\mu = \de_{\rm LT} \, A_\mu$, 
as expected from Eq.\ \rf{transfrel} given that $\de_{\rm LLT} \, A_\mu = 0$.

\subsection{Noether Identities in EC Theory} \label{Noether}

Noether identities are off-shell identities that hold when a theory has a local symmetry~\footnote{Noether had two theorems.
The first was for global symmetries and showed that they give rise to conserved currents, while the second was for local symmetries 
showing that they lead to off-shell identities that must hold~\cite{en18,den11,Traut62}}.
They can be derived for the whole action but also for any term in the action that is itself invariant.

For example, the Einstein-Hilbert term $S_g$ is invariant under Diffs, LLTs, and LTs.
With $\de S_g = 0$ under LTs as given in Eqs.\ \rf{LTvb} and \rf{LTsc} the Noether identity that follows 
from this directly matches the contracted form of the first Bianchi identity in~\rf{ContractedBianchi1}.
The Noether identity that follows from LLTs using~\rf{LLTvb} and~\rf{LLTsc} directly matches the 
contracted form of the second Bianchi identity in \rf{ContractedBianchi2}.
Under Diffs, using \rf{Diffvb} and \rf{Diffsc},
the identity that follows has the form:
\bea
&&(D_\mu - T^\la_{\pt{\la}\la\mu}) G^{\mu\nu} + T_{\la\mu}^{\pt{\la\mu}\nu} G^{\mu\la} 
- \half R^{\al\be\mu\nu} \hat T_{\mu\al\be} 
\nonumber \\
&& \quad\quad\quad\quad\quad
-\half \usc\nu a b  \lvb \al a  \, \lvb \be b \left[ G^{\al\be} - G^{\be\al}  + (D_\si - T^\la_{\pt{\la}\la\si}) \hat T^{\si\al\be} \right]
= 0 \, ,
\label{DiffNoether1}
\eea
where this is identically satisfied off shell by virtue of the identities from LTs and LLTs, 
or equivalently by virtue of the two Bianchi identities.
This gives an illustration of how the three identities under LTs, LLTs, and Diffs are linked
as indicated by Eq.\ \rf{transfrel}.

Similarly, the matter term $S_{g,m} $ is invariant under LTs, LLTs, and Diffs
when the matter fields $f^\ps$ are dynamical and the Lagrangian 
${\cal L}_m (\vb \mu a,\nsc \mu a b, f^\ps)$ is a scalar under each transformation.
For simplicity, the matter fields can be put on shell so that Eq.\ \rf{feq} holds.
In this case, the three identities that follow respectively from LTs, LLTs, and Diffs are:
\beq
(D_\mu - T^\la_{\pt{\la}\la\mu}) T_e^{\pt{e}\mu\nu} + T_{\la\mu}^{\pt{\la\mu}\nu} T_e^{\pt{e}\mu\la} 
+ \half R^{\al\be\mu\nu} S_{\om \, \mu\al\be}  = 0 \, ,
\label{LTNoether2}
\eeq
\beq
T_e^{\pt{e}\mu\nu} - T_e^{\pt{e}\nu\mu}  = (D_\si - T^\la_{\pt{\la}\la\si}) S_{\om \,}^{\pt{\om \,}\si\mu\nu} \, ,
\label{LLTNoether2}
\eeq
\bea
&&(D_\mu - T^\la_{\pt{\la}\la\mu}) T_e^{\pt{e}\mu\nu} + T_{\la\mu}^{\pt{\la\mu}\nu} T_e^{\pt{e}\mu\la} 
+ \half R^{\al\be\mu\nu} S_{\om \, \mu\al\be} 
\nonumber \\
&& \quad\quad\quad\quad\quad
-\half \usc\nu a b  \lvb \al a  \, \lvb \be b \left[ T_e^{\pt{e}\al\be} - T_e^{\pt{e}\be\al}  
- (D_\si - T^\la_{\pt{\la}\la\si}) S_{\om \,}^{\pt{\om \,}\si\al\be} \right]
= 0 \, .
\label{DiffNoether2}
\eea
In this case, there are no geometric Bianchi identities that enforce these
identities as was the case with the Einstein-Hilbert term.
Instead, these identities are the result of the local symmetries,
with the matter fields put on shell.
Note that the gravitational fields are still off shell in these identities,
and only the matter fields have been put on shell.
The fact that the spin connection appears in the identity stemming from Diffs in Eq.\ \rf{DiffNoether2}
appears problematic, since it is not a covariant tensor.
However, the identity for Diffs is a combination of the identities for LTs and LLTs,
as expected from Eq.\ \rf{transfrel}.
Thus, the identity for Diffs is automatically satisfied as a result 
of the identities for LTs and LLTs.

\subsection{Theoretical Consistency and Energy-Momentum Conservation}\label{EMeffconsv}

In GR, in the absence of spin and torsion,
the Einstein equations reduce to $\tilde G^{\mu\nu} = \ka T_e^{\pt{e}\mu\nu}$,
where the tilde denotes that the connection used to define the curvature 
 and Einstein tensor is the Levi-Civita connection $\tilde \Ga^\la_{\pt{\la}\mu\nu}$.
The Einstein tensor in Riemann space is symmetric, obeying $\tilde G^{\mu\nu} = \tilde G^{\nu\mu}$,
and the relevant contracted Bianchi identity is $\tilde D_\mu \tilde G^{\mu\nu} = 0$,
where the covariant derivative, $\tilde D_\mu$, uses $\tilde \Ga^\la_{\pt{\la}\mu\nu}$.
The identities due to LTs and LLTs in Eqs.\ \rf{LTNoether2} and \rf{LLTNoether2}
with zero torsion and putting the matter fields on shell reduce, 
respectively, to $\tilde D_\mu T_e^{\pt{e}\mu\nu} = 0$
and $T_e^{\pt{e}\mu\nu} = T_e^{\pt{e}\nu\mu} $.
These two results automatically satisfy the identity for Diffs in Eq.\ \rf{DiffNoether2} 
with the torsion set to zero.
Thus in GR there is complete consistency between the Bianchi identities,
the Einstein equations, and the equations of motion for the matter fields.
As a result of these, the energy-momentum tensor 
is both symmetric and covariantly conserved.  

In EC theory,
there is also full consistency between the contracted Bianchi identities,
the Einstein and spin connection equations, 
and the equations of motion for the matter fields.
However, in this case, in the presence of spin density and torsion,
$G^{\mu\nu}$ is not symmetric and $D_\mu G^{\mu\nu} \ne 0$,
and therefore $T_e^{\pt{e}\mu\nu} \ne T_e^{\pt{e}\mu\nu}$ 
and $D_\mu T_e^{\pt{e}\mu\nu}  \ne 0$.
These properties of energy-momentum in EC theory have been examined 
and discussed previously~\cite{sciama62,Hehl74,akp74,Hehl76}.
(See also~\cite{HehlRMP76,at06,PBO17}). 
In particular, with spin density and torsion present and on shell,
it has been shown that an effective Riemann geometry
with a conserved and symmetric energy-momentum tensor
can be identified.

To see this, first use that the curvature tensor $R^\ka_{\pt{\ka}\la\mu\nu}$ 
can be separated into a Riemann part $\tilde R^\ka_{\pt{\ka}\la\mu\nu}$ 
and additional terms involving the contorsion:
\beq
R^\ka_{\pt{\ka}\la\mu\nu} = \tilde R^\ka_{\pt{\ka}\la\mu\nu}
+ [ \tilde D_\mu K^\ka_{\pt{\ka}\nu\la} + K^\ka_{\pt{\ka}\mu\si} K^\si_{\pt{\si}\nu\la}
- ( \mu \leftrightarrow \nu ) ] \, .
\label{Rcurvetilde}
\eeq
The torsion can then be eliminated on shell in terms of
the spin density using Eq.\ \rf{TorsionEq},
which allows the on-shell contorsion tensor to be written as 
\beq
K^{\la\mu\nu} = - \fr \ka 2 (S_{\om  }^{\pt{\om \, } \la\mu\nu} 
- S_{\om  }^{\pt{\om \, } \mu\nu\la}  - S_{\om  }^{\pt{\om \, } \nu\mu\la} 
+ g^{\mu\nu} S_{\om  }^{\pt{\om \,} \la} - g^{\la\mu} S_{\om  }^{\pt{\om \,} \nu} ) \, ,
\label{Kspin}
\eeq
where, $S_{\om  }^{\pt{\om \,} \mu} = S_{\om \, \pt{\si} \si }^{\pt{\om \,} \si \pt{\si} \mu}$.
With this substituted into Eq.\ \rf{Rcurvetilde},
an effective theory can be found that uses the Riemann curvature.
However, there is still the caveat that minimal couplings in the full theory 
depend on the Cartan connection $\Ga^\la_{\pt{\la}\mu\nu}$,
not $\tilde \Ga^\la_{\pt{\la}\mu\nu}$, so the full theory remains non-Riemann.
Nonetheless, with the torsion eliminated in this way on shell,
the Einstein equation in \rf{EinsteinEq} becomes
\bea
\tilde G^{\mu\nu} = \ka T_{\rm eff}^{\mu\nu}
+ \fr {\ka^2} 4 (2 S_{\om  }^{\pt{\om}\mu\al\be}  S_{\om \, \al\be}^{\pt{\om \, a\be} \nu} 
+ 2 S_{\om \, }^{\pt{\om \,}\mu\nu\si} S_{\om \, \si } 
- S_{\om \, }^{\pt{\om \,}\mu\al\be} S_{\om \, \, \al\be}^{\pt{\om } \nu} )
\nonumber \\
- \fr {\ka^2} 8 g^{\mu\nu} (S_{\om  }^{\pt{\om}\al\be\ga} S_{\om \, \ga\al\be} 
- S_{\om  }^{\pt{\om}\al\be\ga} S_{\om \, \al\be\ga}  - S_{\om \, \si} S_{\om  }^{\pt{\om \,} \si}  ) \, ,
\label{GtildeSw}
\eea
where an effective energy-momentum tensor is defined as
\beq
T_{\rm eff}^{\mu\nu} = T_e^{\mu\nu} 
- \half \tilde D_\si (S_{\om  }^{\pt{\om}\si\mu\nu} + S_{\om  }^{\pt{\om}\mu\nu\si} + S_{\om  }^{\pt{\om}\nu\mu\si} ) \, .
\label{Teff}
\eeq
Note that $T_{\rm eff}^{\mu\nu}$ has the form of a Belinfante-Rosenfeld 
energy-momentum tensor~\cite{Hehl74,akp74,Hehl76},
where in a theory with spin the extra added spin-density terms 
lead to a redefined energy-momentum tensor that is symmetric.
Although Eq.\ \rf{GtildeSw} is not yet symmetric,
the LLT Noether identity in Eq.\ \rf{LLTNoether2} 
rewritten in terms of $\tilde D_\si$ and additional quadratic terms in 
the spin density can be used
to eliminate the antisymmetric part of the effective energy momentum
from Eq.\ \rf{GtildeSw}.
This leaves an Einstein equation that is effectively Riemann:
\bea
\tilde G^{\mu\nu} = \ka T_{\rm eff}^{(\mu\nu)}
+ \fr {\ka^2} 2 [ (S_{\om  }^{\pt{\om}\mu\al\be}  S_{\om \, \al\be}^{\pt{ \om \, \al\be} \nu}
+ S_{\om }^{\pt{\om} \nu\al\be}  S_{\om \, \al\be}^{\pt{ \om \, \al\be} \mu})
+ ( S_{\om \, \pt{\om } \si}^{\pt{\om \,} \mu\nu} + S_{\om \, \pt{\om }\, \si}^{\pt{\om \,} \nu\mu}) S_{\om  }^{\pt{\om \,} \si}  ]
\nonumber \\
- \fr {\ka^2} 8 g^{\mu\nu} [ (S_{\om  }^{\pt{\om}\al\be\ga} S_{\om \, \ga\al\be} 
- S_{\om  }^{\pt{\om}\al\be\ga} S_{\om \, \al\be\ga}  - S_{\om \, \si} S_{\om  }^{\pt{\om \,} \si}  ) \, ,
\quad\quad\,
\label{GtildeSw2}
\eea
where $T_{\rm eff}^{(\mu\nu)} = \half (T_{\rm eff}^{\mu\nu}  + T_{\rm eff}^{\nu\mu}) $.
In this form, the right-hand side is symmetric and
is consistent with $\tilde G^{\mu\nu} = \tilde G^{\nu\mu}$,
and since $\tilde D_\mu \tilde G^{\mu\nu} = 0$,
the combined terms on the right-hand side also have a 
vanishing covariant divergence with respect to $\tilde D_\mu$.
Thus, in this context, the spin density acts effectively like additional
contributions to the energy-momentum.

Since torsion is extremely weak,
and since the coupling $\ka$ is small,
the quadratic contributions $\sim \ka^2 S_{\om  }^2$ can
be neglected at leading order in a perturbative treatment.
In this approximation, 
\beq
T_{\rm eff}^{\mu\nu} \simeq T_e^{(\mu\nu)} 
- \half \tilde D_\si (S_{\om  }^{\pt{\om}\mu\nu\si} + S_{\om  }^{\pt{\om}\nu\mu\si} ) \, ,
\label{Teffapprox}
\eeq
is symmetric,
and the effective Einstein equation reduces to $\tilde G^{\mu\nu} \simeq \ka T_{\rm eff}^{\mu\nu}$.
From this it follows that $\tilde D_\mu T_{\rm eff}^{\mu\nu} \simeq 0$,
and in a flat spacetime limit with negligible torsion,
$\partial_\mu T_{\rm eff}^{\mu\nu} \simeq 0$.

%%%%%%%%%%%%%%%%%%%%%%%%%%%%%%%%%%%%%%%%%%%%%%%%%%%%%%

\section{Spacetime Symmetry Breaking} \label{sec3}

Many ideas have been put forward for how spacetime symmetries might be broken,
including mechanisms in string theory, above threshold cosmic rays, modified gravity, noncommutative geometry, 
loop quantum gravity, spacetime foam, Chern Simons gravity, Ho\v rava gravity, and massive gravity.
See, for examples,~\cite{akss1,akss2,akss3,scsg1997,chklo2001,jdb04,GAC2013,rgjp99,bmst2005,frkcr2004,sbfrk2004,
aclt07,rjsp,kgfk2018,sl2013,sh2014,lh2019,ayp2020,ph09,Muk10,Sot11,Wang17,mg1,mg2}.

However, in EC theory working at the level of observer-independent effective field theory,
as in the framework of the SME,
violation of spacetime symmetry
occurs when a fixed background tensor field interacts with a dynamical gravitational 
or matter field~\cite{akgrav04,rb15a,rb15b,rbas16,rbSym17,rbhbyw19,rbyySym21,akzl21a,akzl21b}.
In general, the backgrounds can have components with
respect to both spacetime frames and local tangent spaces.
The symmetry breaking occurs because the background fields 
are fixed and do not transform under Diffs, LLTs, or LTs,
while fully dynamical fields do transform.
The combination of an active transformation for fully dynamical fields with the
background fields being held fixed is referred to as a particle transformation.
However, at the same time, a physical theory must be observer independent,
which requires invariance under general coordinate transformations
and passive changes of basis states in local tangent spaces.
These are called observer transformations, 
and under them the components
of fixed backgrounds transform passively,
and the action $S$ is mathematically left unchanged.
In EC theory with no fixed backgrounds,
mathematical observer transformations of
Diffs, LLTs, and LTs can be written in the form
of inverse transformations to the corresponding active transformations.
However, when fixed background fields are present,
the particle transformations are broken,
while the mathematical observer invariances must still hold.

\subsection{EC Theory with Background Fields}

The generic form for the action of an EC theory with fixed background fields can be written as
\bea
&& S = S_g + S_{\bar k_X} + S_{g, \bar k_Y}  + S_{g,m , \bar k_Z} 
\nonumber \\
&& \quad = \fr 1 {2 \ka} \int d^4x \, e \, R (\vb \mu a,\nsc \mu a b) 
- \fr 1 {2 \ka} \int d^4x \, e \, {\cal U} (\vb \mu a, \bar k_X) 
\nonumber \\
&&\quad\quad 
+  \fr 1 {2 \ka} \int d^4x \, e \, {\cal L}_{g, \bar k_Y}  (\vb \mu a,\nsc \mu a b, \bar k_Y)
+  \int d^4x \, e \, {\cal L}_{g, m, \bar k_Z}  (\vb \mu a,\nsc \mu a b, f^\ps, \bar k_Z) \, ,
\label{kECS}
\eea
where the total action has been divided into four terms.
The first term, $S_g $, has the Einstein-Hilbert form,
which depends on the dynamical vierbein and spin connection.
The second term couples the vierbein directly
to fixed backgrounds denoted generically as $\bar k_X$,
where the bar indicates that $\bar k_X$ is a fixed field,
and the indices $X$ generically label the 
spacetime and local indices carried by it.
The remaining two terms contain fixed background fields,
which are denoted generically as $\bar k_Y$ and $\bar k_Z$,
where $Y$ and $Z$ generically label the 
spacetime and local indices carried by each background.
In the last term, $f^\ps$ generically denotes all the dynamical matter fields.
It is assumed that the three backgrounds $\bar k_X$, $\bar k_Y$, or $\bar k_Z$
are different from each other and that no covariant derivatives 
act directly on them.
However, if symmetry transformations or integrations by parts 
are performed in the action,
this can result in expressions where covariant derivatives act on the backgrounds,
in which case it is assumed that $\bar k_X$, $\bar k_Y$, or $\bar k_Z$ 
are not covariantly constant.

The terms $S_{\bar k_X}$, $S_{g, \bar k_Y}$, and $S_{g,m, \bar k_Z}$ 
are all symmetry-breaking terms.  
$S_{\bar k_X}$ is a potential term,
which includes possible mass terms for the vierbein formed using background fields.
There is no dependence on the spin connection in $S_{\bar k_X}$ ,
since that would have to originate from
covariant derivatives acting on the backgrounds $\bar k_X$.
$S_{g, \bar k_Y}$ is a pure-gravity term,
consisting of interactions between the vierbein, spin connection
and the backgrounds $\bar k_Y$.
To maintain covariance,
it is assumed that any terms in $S_{g, \bar k_Y}$ consist only of 
couplings between the backgrounds $\bar k_Y$ with the curvature, torsion,
or covariant derivatives of the curvature or torsion.
Since both $S_{\bar k_X}$ and $S_{g, \bar k_Y}$exclude matter contributions,
they both carry the dimensional coupling, $1/2\ka$,
as in the Einstein-Hilbert term.
The last term, $S_{g,m , \bar k_Z} $,
is a matter-gravity term,
consisting of interactions between the vierbein, spin connection,
matter fields, and the backgrounds $\bar k_Z$.
It also contains conventional matter-gravity couplings that
do not couple to the backgrounds $\bar k_Z$.

In general, there is some ambiguity between how the backgrounds
$\bar k_X$, $\bar k_Y$ and $\bar k_Z$ are determined,
since not all of the background coefficients 
are independent or physical.
This is because coordinate changes and field redefinitions
can be used to move sensitivity to spacetime symmetry breaking
from one sector to another,
including between the potential, pure-gravity, and matter-gravity 
sectors~\cite{akgrav04,akjt2,yb15,dcpm02,rbhbyw19}.
With this in mind, it should be assumed that before splitting the
full action in~\rf{kECS} into these sectors
that any unphysical coefficients have been removed and that
field redefinitions and coordinate choices have been made that fix an 
observable set of backgrounds $\bar k_X$, $\bar k_Y$ and $\bar k_Z$.
Note that in some cases, an observable set of backgrounds might
consist of combinations of backgrounds with couplings to different 
particle species~\cite{akjt2}.

Under particle Diffs, LLTs, and LTs,
all three of the background fields remain fixed, transforming as
\beq
\bar k_X \rightarrow \bar k_X \, , \quad\quad
\bar k_Y \rightarrow \bar k_Y \, , \quad\quad
\bar k_Z \rightarrow \bar k_Z \, ,
\label{ktransf}
\eeq
while under observer Diffs, LLTs, and LTs,
they transform passively according to their representations
as spacetime or local tensors as indicated by the index labels $X$, $Y$, and $Z$.
Since the backgrounds $\bar k_X$, $\bar k_Y$ and $\bar k_Z$ are assumed to
be observable, the three symmetry-breaking sectors are distinct and independent.  
Thus, the potential, ${\cal U} (\vb \mu a, \bar k_X)$,
and the Lagrangians, ${\cal L}_{g, \bar k_Y}  (\vb \mu a,\nsc \mu a b, \bar k_Y) $
and ${\cal L}_{g, m, \bar k_Z}  (\vb \mu a,\nsc \mu a b, f^\ps, \bar k_Z) $
must each be a scalar under observer spacetime transformations to maintain
observer independence.
However, they are not scalars under the broken particle transformations.

\subsection{Explicit versus Spontaneous Breaking}

To understand the nature of the background fields $\bar k_X$, $\bar k_Y$, and $\bar k_Z$,
as well as their effects in the context of effective field theory, 
a distinction must be made between when the spacetime symmetry breaking is 
spontaneous versus when it is explicit.

With spontaneous breaking, the symmetry is hidden.
It is only the vacuum solution that breaks the symmetry,
while the full solutions, including excitations in the
form of NG modes and additional massive Higgs-like modes
are fully dynamical and transform appropriately under 
all spacetime transformations.
For example, 
with spontaneous breaking, 
the backgrounds $\bar k_X$, $\bar k_Y$, and $\bar k_Z$ all equal vacuum expectation values,
\beq
\bar k_X = \vev{K_X} \, , \quad\quad \bar k_Y = \vev{K_Y} \, , \quad\quad \bar k_Z = \vev{K_Z} \, ,
\label{Kvev}
\eeq 
where $K_X$ , $K_Y$, and $K_Z$ are dynamical fields.
As dynamical fields,
all three sets of components $K_X$, $K_Y$,  and $K_Z$ 
undergo field variations and have equations of motion.
They are very much like any other dynamical field components
except that they have vacuum values 
$\vev{K_X}$, $\vev{K_Y}$, and $\vev{K_Z}$,
which spontaneously break spacetime symmetries.
In contrast, with explicit breaking, 
the backgrounds $\bar k_X$, $\bar k_Y$, and $\bar k_Z$
are fixed nondynamical fields that are inserted into the Lagrangian.
They have no dynamical field variations
and therefore no equations of motion.

For simplicity,
consider the case of a vector background field,
which has components $ \bar b_\mu$ with respect to the spacetime frame
and components $ \bar b_a$ with respect to a local basis.
Since the vector is fixed under particle transformations,
its components are unchanged in either frame, 
obeying $\de  \bar b_\mu = 0$ and $\de  \bar b_a = 0$ 
under particle Diffs, LLTs, and LTs.

In the case of spontaneous breaking,
$ \bar b_\mu$ and $ \bar b_a$ are vacuum values of 
a dynamical field with components $B_\mu$ or $B_a$,
\beq
\vev{B_\mu} = \bar b_\mu \, , \quad \vev{B_a} = \bar b_a \, .
\label{Bvevs}
\eeq
The dynamical field components are related to each other by the vierbein 
as $B_\mu = \vb \mu a B_a$.
With spontaneous breaking, the vierbein also has a vacuum value,
$\vev{\vb \mu a}$,
which in a Minkowski background in Cartesian coordinates is
$\vev{\vb \mu a} = \de^{\pt{\mu} a}_\mu$.
The vierbein vacuum value relates the two vector vacuum values as
\beq
\bar b_\mu =  \vev{\vb \mu a} \bar b_a \, .
\label{bmuveveba}
\eeq
Since the vacuum solutions, $\bar b_\mu$, $\bar b_a $, and $\vev{\vb \mu a}$
carry both spacetime and local indices,
all three of the symmetries Diffs, LLTs, and LTs are spontaneously broken.
In particular, the existence of a vacuum geometry with $\vev{\vb \mu a} \ne 0$
requires that both $\bar b_\mu$ and $\bar b_a$ must be nonzero
if either of them is.
 
With spontaneous breaking, any action term, ${\cal L} = J^\mu B_\mu$, 
involving a current $J^\mu$ interacting 
with the dynamical field $B_\mu$, 
can also be written in terms of local components,
since $J^\mu B_\mu = J^a B_a$.
Hence, as the vector field separates into a vacuum solution and excitations,
any terms in the effective action of the form $J^\mu \bar b_\mu$ or $J^a \bar b_a$
are physically linked by the vacuum vierbein $\vev{\vb \mu a}$ and its excitations.  
Similarly, couplings to $\bar b^\mu$, with an upper index, are linked to $\bar b_\mu$, 
since these are related by the vacuum solution for the metric,
$\vev{g_{\mu\nu}} = \vev{\vb \mu a} \vev{\vb \nu b} \et_{ab}$.
In local frames, components $\bar b_a$ and $\bar b^a = \et^{ab} \bar b_b$ are 
directly related by the Minkowski metric.

With explicit breaking, there are no vacuum values of physical fields.
There are only fixed nondynamical backgrounds.  
A fixed nonzero background vector must have nonzero components
$\bar b_\mu$ in the spacetime frame and components $\bar b_a$ in the local basis.
However, there is no linkage between these given by a physical or vacuum vierbein.
Similarly, background components $\bar b^\mu$ are not linked to $\bar b_\mu$
by the physical metric.
However, in local frames $\et_{ab}$ is the metric,
so components $\bar b_a$ and $\bar b^a$ are related using it.  

Any Lagrangian terms such as $J^\mu \bar b_\mu$, $J_\mu \bar b^\mu$,
and $J^a \bar b_a$ are therefore all physically distinct when the
symmetry breaking is explicit.
Taken separately, each of the backgrounds $\bar b_\mu$,
$\bar b^\mu$, or $\bar b_a$ might have couplings that explicitly break 
one or more of the spacetime symmetries, Diffs, LLTs, or LTs,
but not necessarily any two or more of them at the same time.
This generalizes as well to tensors with more than one index,
including possibly tensors that have both spacetime and local indices.
An extensive list of examples of possible distinct tensor backgrounds,
including which spacetime symmetries they explicitly break, 
is given in~\cite{akzl21a},
and examples of their phenomenological implications are explored in~\cite{akzl21b}.

Regardless of whether the symmetry breaking is spontaneous or explicit,
the action must be a scalar under observer Diffs, LLTs, and LTs.
In this case, background vector components, $\bar b_\mu$,
$\bar b^\mu$, or $\bar b_a$, have conventional observer transformations
appropriate for the type of index they carry.  
For example,
the transformations of $\bar b_a$ under observer Diffs, LLTs, and LTs,
are given, respectively, 
as\footnote{Technically, here, the parameters $\ep^a$ and $\ep_a^{\pt{a} b}$
for observer transformations have the opposite signs of those used in~\rf{Ab}
and~\rf{Amu}; however, they can simply be redefined with a minus sign
so that the transformations have the same mathematical form.}
\bea
\de_{\rm Diff} \, \bar b_a = \xi^\nu \partial_\nu \bar b_a \, , \quad\quad
\nonumber \\
\de_{\rm LLT} \, \bar b_a = - \ep_a^{\pt{a} c} \bar b_c \, , \quad\quad
\nonumber \\
\de_{\rm LT} \, \bar b_a = \ep^c \ivb \mu c D^{(\om)}_{\, \mu} \bar b_a \, ,
\label{batrans}
\eea
while $\bar b_\mu $ transforms under
observer Diffs, LLTs, and LTs as
\bea
\de_{\rm Diff} \, \bar b_\mu =  (\partial_\mu \xi^\nu) \bar b_\nu 
+ \xi^\nu \partial_\nu \bar b_\mu \, , 
\quad\quad\quad\quad\quad\quad\quad\quad\quad
\nonumber \\
\de_{\rm LLT} \, \bar b_\mu = 0 \, , \quad\quad\quad\quad\quad\,\,
\quad\quad\quad\quad\quad\quad\quad\quad\quad\quad\quad
\nonumber \\
\de_{\rm LT} \, \bar b_\mu = (D^{(\om)}_{\, \mu} \ep^b) \ivb \nu b \bar b_\nu
+ \ep^b \ivb \nu b D_{\nu} \bar b_\mu + \ep^b T^\nu_{\pt{a} b \mu} \bar b_\nu\, . \,\,
\label{bmutrans}
\eea

\subsection{Equations of Motion}

Before writing the variations of the full action $S$,
it is convenient to make some definitions regarding 
the separate variations of the potential, pure-gravity and 
matter-gravity terms.

First, the variation of the potential term $S_{\bar k_X}$
with respect to the vierbein can be written as
\beq
\de S_{\bar k_X} = \int d^4x \, e \, (- \fr 1 {2 \ka} \bar {\cal U}^{\mu\nu} \lvb \nu a) \, \de \vb \mu a \, ,
\label{ECSX}
\eeq
where $\bar {\cal U}^{\mu\nu}$ is written with a bar over it to denote that it
includes contributions from the backgrounds $\bar k^X$.
A coupling $1/2\ka$ is included in $S_{\bar k_X}$,
since it does not include matter fields.

Next,
define the variations of $S_{g, \bar k_Y}$ with respect to
the vierbein and spin connection as:
\beq
\de S_{g, \bar k_Y} = \int d^4x \, e \, 
\left[ -\fr 1 {\ka} \bar {\cal G}^{\mu\nu} e_{\nu a} \, \de \vb \mu a 
+  \fr 1 {2 \ka} \bar {\cal T}^{\mu\al\be} e_{\al a} e_{\be b}  \de \nsc \mu a b \right] \, .
\label{ECdeltaSk}
\eeq
Here, $\bar {\cal G}^{\mu\nu}$ and $\bar {\cal T}^{\mu\al\be}$
are written with bars over them to indicate that they include 
contributions coming from the background fields $\bar k_Y$.
Since $S_{g, \bar k_Y}$ contains a factor of $1/2\ka$,
defining $\bar {\cal G}^{\mu\nu}$ and $\bar {\cal T}^{\mu\al\be}$ in this way
indicates that these terms have similar mass dimensions as the curvature and torsion
terms arising from the Einstein-Hilbert term.

Lastly, the variations of $S_{g,m , \bar k_Z}$ with respect to
the vierbein, spin connection, and matter fields 
are written as:
\beq
\de S_{g,m, \bar k_Z} = \int d^4x \, e \, \left[ \bar T_e^{\mu\nu} e_{\nu a} \, \de \vb \mu a 
+  \half \bar S_{\om \,}^{\pt{\om \, }\mu\al\be} e_{\al a} e_{\be b} \de \nsc \mu a b 
+ \fr {\de S_{g, m, \bar k_Z}} {\de f^\ps} \de f^\ps \right] \, .
\label{ECdeltaSmk}
\eeq
Here, the energy-momentum tensor, $\bar T_e^{\mu\nu}$, 
and spin density tensor, $\bar S_{\om \, \pt{\mu} a b}^{\pt{\om \, }\mu}$, 
are written with bars over them to indicate they include 
contributions coming from both matter and the background fields $\bar k_Z$.
If the backgrounds $\bar k_Z$ vanish, 
$\bar T_e^{\mu\nu}$ and $\bar S_{\om \, \pt{\mu} a b}^{\pt{\om \, }\mu}$
reduce to the energy-momentum and spin density for the matter fields alone,
which can then be written without using bars.

With these definitions, the variation of the full action with respect to the dynamical fields,
$\vb \mu a $, $\nsc \mu a b$, and $f^\ps$,
has the form
\bea
&&\de S = \int d^4x \, e \, 
\left[ [-\fr 1 {\ka} (G^{\mu\nu} + \bar {\cal U}^{\mu\nu} + \bar {\cal G}^{\mu\nu}) 
+ \bar T_e^{\mu\nu}] e_{\nu a} \de \vb \mu a \right.
\nonumber \\
&& \quad\quad+ \left. [\fr 1 {2 \ka} (\hat T^{\mu\al\be} + \bar {\cal T}^{\mu\al\be})  
+ \half \bar S_{\om \,}^{\pt{\om \, }\mu\al\be} ] e_{\al a} e_{\be b} \, \de \nsc \mu a b
+ \fr {\de S_{g,m, \bar k_Z}} {\de f^\ps} \de f^\ps \right] \, .
\,
\label{ECdeltaSkbar}
\eea
Setting $\de S = 0$ for dynamical variations $\de \vb \mu a $, $\de \nsc \mu a b$, and $\de f^\ps$
gives the equations of motion for the vierbein,
spin connection, and matter fields, 
respectively, as
\bea
G^{\mu\nu} + \bar {\cal U}^{\mu\nu} + \bar {\cal G}^{\mu\nu}
= \ka \bar T_e^{\mu\nu} \, ,  \quad\quad
\label{EinsteinEqkbar} \\
\hat T^{\la\mu\nu}  +  \bar {\cal T}^{\la\mu\nu}
=   - \ka \bar S_{\om \, \pt{\mu} }^{\pt{\om \, } \la\mu\nu} \, ,
\label{TorsionEqkbar} \\
\fr {\de S_{g, m, \bar k_Z}} {\de f^\ps} = 0 \, .
\quad\quad\quad\,\,\,\,
\label{feqkbar}
\eea
In these equations,
the quantities $ \bar {\cal U}^{\mu\nu}$ and 
$ \bar {\cal G}^{\mu\nu}$ 
can be interpreted in two different ways.
In the first, they act, respectively, as corrections to the curvature,
which depend on the backgrounds $\bar k_X$ and $\bar k_Y$.
Alternatively, they can be interpreted as belonging
on the right-hand side of~\rf{EinsteinEqkbar},
where in that case they contribute to the energy-momentum.
Similarly, the quantity $\bar {\cal T}^{\la\mu\nu}$ can
be interpreted as corrections to the torsion,
which depend on the backgrounds $\bar k_Y$,
or they can go on the right-hand side of~\rf{TorsionEqkbar}
and act as contributing to the spin density.
However, with these quantities on the right-hand sides, 
the coupling $\ka$ does not appear when 
$\bar k_X$ and $\bar k_Y$ interact with the vierbein and 
spin connection as it does when matter fields and $\bar k_Z$ couple to them.
For this reason, it is more natural to keep the quantities 
$ \bar {\cal U}^{\mu\nu}$, $ \bar {\cal G}^{\mu\nu}$ 
and $\bar {\cal T}^{\la\mu\nu}$ on the left-hand sides of
the equations of motion.

\subsection{No-go Results and Noether Identities}

The issue of whether no-go conditions apply when 
local spacetime symmetry breaking occurs
can be examined using Noether identities that hold 
as a result of observer independence.
While the background fields break particle spacetime symmetries,
the mathematical observer symmetries in the action must still hold
so that observer independence is maintained.
Thus, the observer transformations can be used to find
Noether identities even when background fields are present.
Since observer Diffs, LLTs, and LTs are related,
it suffices to consider the Noether identities resulting
from only LTs and LLTs.
The action terms $S_g$, $S_{\bar k^X}$, $S_{g, \bar k_Y}$, and $S_{g, m, \bar k_Z}$
are each separately unchanged under observer LTs and LLTs,
and Noether identities can be found from each one and
for each symmetry.

For the Einstein-Hilbert term, setting $\de S_g = 0$ 
gives
\beq
\de S_g  = \int d^4x \, e \, 
\left[ -\fr 1 {\ka} G^{\mu\nu} e_{\nu a} \de \vb \mu a 
+ \fr 1 {2 \ka} \hat T^{\mu\al\be} e_{\al a} e_{\be b} \, \de \nsc \mu a b \right] = 0 \, ,
\label{SgObsvVarsEC}
\eeq
where $\de \vb \mu a $ and $\de \nsc \mu a b$ are variations
given in~\rf{LTvb} and~\rf{LTsc}
for LTs and~\rf{LLTvb} and~\rf{LLTsc} for LLTs.
The Noether identities that follow from these
are the same as the contracted forms of the Bianchi identities 
in~\rf{ContractedBianchi1} and~\rf{ContractedBianchi2}.

The action terms $S_{\bar k^X}$, $S_{g, \bar k_Y}$, and $S_{g, m, \bar k_Z}$
are each unchanged as well under observer spacetime transformations.
The variations are given as
\beq
\de S_{\bar k^X} = \int d^4x \, e \, 
\left[ -\fr 1 {\ka}  \bar {\cal U}^{\mu\nu}  \de \vb \mu a 
+ \fr {\de S_{\bar k_X}} {\de \bar k_X} \de \bar k_X \right] =0 \, ,
\label{SXObsvVarsEC}
\eeq
\beq
\de S_{g, \bar k_Y}  = \int d^4x \, e \, 
\left[ [-\fr 1 {\ka}  \bar {\cal G}^{\mu\nu}  \de \vb \mu a  
+ \fr 1 {2 \ka}  \bar {\cal T}^{\mu\al\be} \, \de \nsc \mu a b
+ \fr {\de S_{g, \bar k_Y}} {\de \bar k_Y} \de \bar k_Y \right] =0 \, ,
\label{SYObsvVarsEC}
\eeq
\beq
\de S_{g, m, \bar k_Z} = \int d^4x \, e \, \left[
\bar T_e^{\mu\nu} e_{\nu a} \de \vb \mu a 
+ \half \bar S_{\om \,}^{\pt{\om \, }\mu\al\be} e_{\al a} e_{\be b} \, \de \nsc \mu a b
+ \fr {\de S_{g, m, \bar k_Z}} {\de \bar k_Z} \de \bar k_Z 
+ \fr {\de S_{g,m, \bar k_Z}} {\de f^\ps} \de f^\ps\right] = 0  \, ,
\label{SZObsvVarsEC}
\eeq
where $\de \vb \mu a$, $\de \nsc \mu a b$, $\de \bar k_X$, $\de \bar k_Y$, $\de \bar k_Z$, and $\de f^\ps $
are variations of these fields under observer LTs or observer LLTs.

Notice that variations of the background fields $\bar k_X$, $\bar k_Y$, and $\bar k_Z$ are included 
in these expressions because these fields transform under observer transformations.  
However, they do not appear in the dynamical variations of the full action, $\de S$, 
in~\rf{ECdeltaSkbar}.
Thus, when the Bianchi identities are combined with the 
results that $\de S_{\bar k^X} = \de S_{g, \bar k_Y} = \de S_{g, m, \bar k_Z} = 0$
under observer LTs and LLTs,
consistency with the dynamical equations of motion only holds if
\beq
\int d^4x \, e \fr {\de S_{\bar k_X}} {\de \bar k_X} \de \bar k_X = 0 \, , \quad\,
\int d^4x \, e \fr {\de S_{g, \bar k_Y}} {\de \bar k_Y} \de \bar k_Y  = 0  \, , \quad\,
\int d^4x \, e  \fr {\de S_{g, m, \bar k_Z}} {\de \bar k_Z} \de \bar k_Z = 0 \, .
\label{nogo1}
\eeq

When the integrals in~\rf{nogo1} all vanish under observer LTs,
and when the matter fields, $ f^\ps$, are on shell, obeying~\rf{feqkbar},
the Noether identities that follow from LTs are
\beq
(D_\mu - T^\la_{\pt{\la}\la\mu})  \bar {\cal U}^{\mu\nu} 
+ T_{\la\mu}^{\pt{\la\mu}\nu}  \bar {\cal U}^{\mu\la}  = 0 \, ,
\label{LTNoether2obsU}
\eeq
\beq
(D_\mu - T^\la_{\pt{\la}\la\mu})  \bar {\cal G}^{\mu\nu} 
+ T_{\la\mu}^{\pt{\la\mu}\nu}  \bar {\cal G}^{\mu\la}  
+ \half R^{\al\be\mu\nu}  \bar {\cal T}_{\mu\al\be}  = 0 \, ,
\label{LTNoether2obsG}
\eeq
\beq
(D_\mu - T^\la_{\pt{\la}\la\mu}) \bar T_e^{\pt{e}\mu\nu} 
+ T_{\la\mu}^{\pt{\la\mu}\nu} \bar T_e^{\pt{e}\mu\la} 
+ \half R^{\al\be\mu\nu}  \bar S_{\om \, \mu\al\be} = 0 \, .
\label{LTNoether2obsTe}
\eeq
Similarly, the identities that follow when observer LLTs are made,
and the conditions in~\rf{nogo1} hold with $f^\ps$ on shell, are
\beq
\bar {\cal U}^{\mu\nu} - \bar {\cal U}^{\nu\mu} 
= 0 \, .
\label{LLTNoether2obsU}
\eeq
\beq
\bar {\cal G}^{\mu\nu} - \bar {\cal G}^{\nu\mu} 
= (D_\si - T^\la_{\pt{\la}\la\si})  \bar {\cal T}^{\si\mu\nu} = 0  \, .
\label{LLTNoether2obsG}
\eeq
\beq
\bar T_e^{\pt{e}\mu\nu} - \bar T_e^{\pt{e}\nu\mu} 
= (D_\si - T^\la_{\pt{\la}\la\si})  \bar S_{\om \,}^{\pt{\om \,} \si\mu\nu} = 0 \, .
\label{LLTNoether2obsTe}
\eeq
Comparing these with the contracted Bianchi identities 
in~\rf{ContractedBianchi1} and~\rf{ContractedBianchi2}
and using the equations of motion 
in~\rf{EinsteinEqkbar} and~\rf{TorsionEqkbar},
confirms that these are all compatible as long as
the conditions in~\rf{nogo1} hold.
However, if the integrals in~\rf{nogo1} do not vanish, 
a no-go result follows, and the theory is inconsistent~\cite{akgrav04}.

%%%%%%%%%%%%%%%%%%%%%%%%%%%%%%%%%%%%%%%%%%%%%%%%%%%%%%

\section{Explicit Breaking} \label{sec4}

With explicit breaking,
the backgrounds $\bar k_X$, $\bar k_Y$, and $\bar k_Z$ are nondynamical
and do not satisfy Euler-Lagrange equations,
\beq
\fr {\de S_{\bar k_X}} {\de \bar k_X} \ne 0 \, , \quad\quad
\fr {\de S_{g, \bar k_Y}} {\de \bar k_Y} \ne 0 \, , \quad\quad 
\fr {\de S_{g,m, \bar k_Z}} {\de \bar k_Z} \ne 0 \, ,
\label{kbareq}
\eeq
which makes satisfying the conditions in~\rf{nogo1} problematic.
To examine this, it is useful to rewrite the expression in~\rf{nogo1} in terms of currents,
defining
\beq
J^X =  \fr {\de S_{g, \bar k_X}} {\de \bar k_X} \, ,
\quad\quad
J^Y =  \fr {\de S_{g, m, \bar k_Y}} {\de \bar k_Y} \, ,
\quad\quad
J^Z =  \fr {\de S_{g, \bar k_Z}} {\de \bar k_Z} \, ,
\label{JXY}
\eeq
for the potential, pure-gravity, and matter-gravity sectors, respectively.
The conditions in~\rf{nogo1} can then be written as
\beq
\int d^4x \, e \, J^X \de \bar k_X = 0 \, , \quad\,
\int d^4x \, e \, J^Y  \de \bar k_Y  = 0  \, , \quad\,
\int d^4x \, e  \, J^Z  \de \bar k_Z = 0 \, .
\label{nogo2}
\eeq
Under LTs,
the variations ${\de \bar k_X}$, ${\de \bar k_Y}$, and ${\de \bar k_Z}$
each consist of four local translations with parameters $\ep^a$,
while for LLTs,
they each consist of six local Lorentz transformations 
with parameters $\ep^{ab}$.
Thus, using integrations by parts and the fact that the integrands
must vanish for all $\ep^a$ and $\ep^{ab}$,
up to ten conditions can be extracted from 
each of the three conditions in~\rf{nogo2}.
Thus, there are a total of up to 30 conditions that must hold
if explicit breaking occurs in all three sectors.  
However, this exceeds the number of available degrees of freedom
that are available. 

With explicit breaking, there are at most ten additional degrees of
freedom in the vierbein and spin connection due to the
loss of gauge invariance under LTs and LLTs.
These consist of four degrees of freedom that can normally 
be gauged away using LTs plus another six degrees of freedom
that can normally be gauged away using LLTs.
When the symmetries are explicitly broken,
these degrees of freedom can no longer be gauged away,
and hence all ten can potentially become available to satisfy 
some of the conditions in~\rf{nogo2}.

However, with at most ten extra degrees of freedom,
if explicit breaking of LTs and LLTs occurs in all three sectors, 
then the three sets of conditions in~\rf{nogo2}
cannot hold,
and the theory will be inconsistent.
In the case that two or more sectors 
with special values of $\bar k_X$, $\bar k_Y$, and $\bar k_Z$,
break a combined total of ten or fewer of the
symmetries under LTs and LLTs,
then enough degrees of freedom might be available.
However, with generic values of $\bar k_X$, $\bar k_Y$, and $\bar k_Z$, 
as considered here,
it is not possible to evade a no-go result.

Therefore, it is assumed in the remainder of this paper that only one of the
three types of backgrounds $\bar k_X$, $\bar k_Y$, and $\bar k_Z$ can be
nonzero in a theory with explicit breaking.
In this case, there are up to ten identities that must hold as well
as ten additional degrees of freedom in the vierbein,
making it possible in principle to evade the no-go results.

With explicit breaking occurring in only one sector,
each sector can be examined separately,
in which case only one relevant condition in~\rf{nogo2} 
must hold in each case.
There are, however, a number of ways in which these conditions
might still not hold even if there are enough degrees of freedom.
For example, if the extra degrees of freedom resulting from
explicit breaking do not appear in the relevant identity or 
equations of motion,
then a no-go result follows.
Even if the extra degrees of freedom do appear in the relevant identity,
they still need to provide solutions that exist.
Furthermore, if it turns out that solutions only exist when background tensors
satisfy certain constraints,
then the theory is inconsistent unless the backgrounds are defined
from the start as obeying these constraints.
Thus, the only way to tell for sure if a theory is consistent is to examine whether
the no-go conditions for it can be evaded for all possible values of the
background as defined by the theory.

\subsection{Potential Terms}

To examine theories with symmetry breaking in a potential term,
the background $\bar k^X$ is assumed not to vanish 
while both $\bar k^Y$ and $\bar k^Z$ are set to zero.
The equations of motion for the vierbein and spin connection 
in this case are
\bea
G^{\mu\nu} + \bar {\cal U}^{\mu\nu} 
= \ka T_e^{\mu\nu} \, ,  \quad\quad
\label{EinsteinEqkbarU} \\
\hat T^{\la\mu\nu}  
=   - \ka S_{\om \, \pt{\mu} }^{\pt{\om \, } \la\mu\nu} \, . \,
\label{TorsionEqkbarU}
\eea
The torsion equation is unchanged,
since ${\cal U}$ is assumed not to depend on the spin connection.
If the no-go results are evaded,
the Noether identities that hold for the potential term
under observer LTs and LLTs
are given in~\rf{LTNoether2obsU} and~\rf{LLTNoether2obsU},
which have a form that is consistent with the Bianchi identities,
the identities~\rf{LTNoether2} and~\rf{LLTNoether2},
and the equations of motion in~\rf{EinsteinEqkbarU}
and~\rf{TorsionEqkbarU}.

The relevant condition that must hold to evade the no-go result is 
\beq
\int d^4x \, e \, J^X \de \bar k_X = 0 \, ,
\label{JXnogo}
\eeq
where $J^X  = - \fr {\de \bar {\cal U}} {\de \bar k^X}$,
and $ \de \bar k_X $ are the variations of $\bar k_X $ 
under observer LTs and LLTs,
To evade the no-go result, enough of the extra degrees of freedom in
the vierbein must be present in the theory so that solutions of~\rf{JXnogo} exist.  
These conditions have been examined in a number of 
examples with specific types of 
backgrounds~\cite{rbas16,rbhbyw19,rbyySym21,akzl21a,akrp21}, 
and solutions have been found that evade the no-go results.
At the same time, however,
other examples are known that do not evade them.

For example, backgrounds having the form of a symmetric two-tensor,
such as a background metric, or as a background vierbein,
have been widely investigated in theories of massive gravity~\cite{mg1,mg2}.
These are theories that couple the metric or vierbein to a
background and construct a potential ${\cal U}$ containing
a mass term in such a way that the ghost mode that 
typically appears in massive gravity theories is absent.
Models with a spacetime background having the form of
a fixed Minkowski tensor $\et_{\mu\nu}$
or with a fixed vierbein have been explored,
which evade the no-go results.
Additional ansatz solutions in cosmological or 
Schwarzschild spacetimes that are consistent
have been found as well.
In certain cases,
backgrounds with spacetime dependence can have
consistent solutions provided the metric has a specified form.
However, it is not the case that the condition in~\rf{JXnogo}
can be satisfied for generic backgrounds $\bar k_X $,
and hence no-go results hold in most cases.

\subsection{Pure-Gravity Sector}

This section examines explicit breaking in the pure-gravity sector 
due to the appearance of backgrounds $\bar k_Y$
(with $\bar k_X$ and $\bar k_Z$ set to zero).
The equations of motion in this case are
\bea
G^{\mu\nu} + \bar {\cal G}^{\mu\nu}
= \ka \bar T_e^{\mu\nu} \, ,  \quad\quad
\label{EinsteinEqkbarG} \\
\hat T^{\la\mu\nu}  +  \bar {\cal T}^{\la\mu\nu}
=   - \ka \bar S_{\om \, \pt{\mu} }^{\pt{\om \, } \la\mu\nu} \, ,
\label{TorsionEqkbarG} 
\eea
and if the no-go results are evaded the Noether identities for
observer LTs and LLTs are in~\rf{LTNoether2obsG} and~\rf{LLTNoether2obsG},
which have a form that is consistent with the Bianchi identities,
the identities~\rf{LTNoether2} and~\rf{LLTNoether2},
and the equations of motion in~\rf{EinsteinEqkbarG}
and~\rf{TorsionEqkbarG}.

Note that in an approximately flat and torsionless limit,
and assuming the no-go conditions are evaded,
the identities in~\rf{LTNoether2obsG} and~\rf{LLTNoether2obsG} reduce,
respectively, to 
\beq
\partial_\mu \bar {\cal G}^{\mu\nu} \simeq 0 \, ,
\label{dT=0}
\eeq
\beq
 \bar {\cal G}^{\mu\nu} \simeq \bar {\cal G}^{\nu\mu}  \, .
\label{tmunu=tnumu}
\eeq
However, if $\bar k_Y$ has spacetime dependence, 
it becomes problematic for 
$\partial_\mu \bar {\cal G}^{\mu\nu} \simeq 0$ to hold.
One way to avoid such problems is to assume that
the backgrounds $\bar k_Y$ are constant.
This is the assumption made for backgrounds in the gravity sector 
of the SME when they result from spontaneous breaking.
However, with explicit breaking it is harder to justify such an assumption,
since $\bar k_Y$ is a pre-determined quantity.
Nonetheless, it is assumed here that $\bar k_Y$ is constant.
Presumably, if the no-go results cannot be evaded in this case,
they will be even more problematic in the case where 
$\bar k_Y$ has spacetime dependence.

The relevant condition that must hold to evade the no-go result 
in this case is 
\beq
\int d^4x \, e \, J^Y \de \bar k_Y = 0 \, ,
\label{JYnogo}
\eeq
where $ \de \bar k_Y$ are variations under observer LTs and LLTs.
In covariant form,
the pure-gravity term $S_{g, \bar k_Y}$ is assumed to
consist of expressions that couple the backgrounds $\bar k_Y$
with products of the curvature and torsion, 
as well as with covariant derivatives of the curvature or torsion.
Generic examples of possible terms are given in~\cite{akgrav04}.
Since $\bar k_Y$ is presumably small,
the action is assumed to be linear in $\bar k_Y$.

In~\cite{ybcc18a}, a specific example of a pure-gravity term with explicit breaking
is given with Lagrangian
${\cal L}_{g , \bar k_Y} = \ivb \mu a \ivb \nu b \bar k_{cd}^{\pt{cd}ab} R^{cd}_{\pt{cd}\mu\nu}$,
where the background $\bar k_Y$ in this case is
a nondynamical field $\bar k_{cd}^{\pt{cd}ab}$ that
matches a term in the SME.
With $\bar k_{cd}^{\pt{cd}ab} $ included in the action,
the equations of motion for the vierbein and spin connection
both include contributions from the background.
Interestingly, it is observed in~\cite{ybcc18a}
that the nondynamical background $\bar k_{cd}^{\pt{cd}ab}$
can potentially act as a source of torsion even in vacuum.

Another example of a pure-gravity term is
${\cal L}_{g, \bar k_Y} = \bar k^{\la\mu\nu} K_{\la\mu\nu}$,
where in this case $\bar k_Y$ is a background $\bar k^{\la\mu\nu} $,
and $K_{\la\mu\nu}$ is the contorsion tensor.
Varying the action $S_{g , \bar k_Y}$ with respect to $\lsc \la \mu\nu$
gives $\bar {\cal T}^{\la\mu\nu} = - \bar k^{\mu\la\nu}$
in the equation of motion in~\rf{TorsionEqkbar}.
Thus, in regions of space where the matter fields and spin density equal zero,
Eq.\ \rf{TorsionEqkbar} reduces to 
\beq
\hat T^{\la\mu\nu} = - \bar {\cal T}^{\la\mu\nu} = \bar k^{\\mu\la\nu} \, ,
\label{kbark}
\eeq
showing that here too the background can act as a source
of torsion even in vacuum.

\subsubsection{Perturbation Theory}

In examples such as these,
with explicit breaking in the pure-gravity sector,
their theoretical consistency depends on whether
the identities that follow from~\rf{JYnogo} under observer
LTs and LLTs can hold or not.
At the same time,
in the context of an effective field theory that is used to
analyze tests of gravity on Earth and in the solar system,
it typically suffices to use first-order perturbation theory,
since gravity is weak and torsion has never been detected.
In this case, quadratic or higher-order terms in the 
curvature and torsion can be neglected,
and a post-Newtonian framework can be developed.
Such higher-order terms 
would have the added effect of potentially modifying
the number of propagating degrees of freedom in the theory,
which would be a significant departure from EC theory.
For these reasons, only first-order terms in the curvature or torsion
are considered here in the pure-gravity sector,

With a perturbative approach,
both the consistency and the usefulness of the theory 
must be examined in the case of explicit breaking.
Such an investigation was carried out in Riemann space,
with zero torsion, in~\cite{rb15a},
where it was shown that while it may be possible to 
evade no-go results nonperturbatively in the pure-gravity sector,
in a leading-order perturbative treatment the no-go constraints 
can nonetheless render a post-Newtonian framework useless. 
This is because with explicit breaking the extra degrees of 
freedom in the metric or vierbein,
which are normally gauge degrees of freedom,
do not appear in a way that allows the no-go results to be evaded.
This is due to the fact that the linearized curvature is gauge invariant.
Hence, the needed extra degrees of freedom disappear from it.
As a result, conditions must be imposed either on the curvature tensor, 
which limits the geometry, or on the background fields themselves,
which invalidates the premise that they are prescribed quantities.

A similar general argument concerning the consistency and usefulness of
a perturbative approach can be made in Riemann-Cartan space as well.
This is because the linearized curvature and torsion are both gauge invariant
under infinitesimal Diffs, LTS, and LLTs, 
when excitations $(\vb \mu a - \de_\mu^a)$ and $\nsc \mu a b$ are small.
This causes the extra modes in the vierbein and spin connection
due to the gauge breaking to disappear at the linearized level.

To see this, consider a zeroth-order flat background with zero torsion,
where $\vb \mu a = \de^a_\mu$ and $\nsc \mu a b =0$.
With such a background, when the vierbein acts on a field it converts local
indices to spacetime indices,
and hence the distinction between them can be dropped.
Linearizing the curvature in~\rf{Rvieb} and the torsion in~\rf{Torvieb}
in this way gives the first-order expressions:
\beq
R_{\ka\la\mu\nu} \simeq \partial_\mu \om_{\nu\ka\la} - \partial_\nu \om_{\mu\ka\la} \, ,
\label{linR}
\eeq
\beq
T_{\la\mu\nu} \simeq \partial_\mu \lvb \nu\la - \partial_\nu \lvb \mu\la
+ \om_{\mu\la\nu} - \om_{\nu\la\mu} \, ,
\label{linT}
\eeq
The linearized infinitesimal transformations under Diffs, LTs, and LLTs are:
\beq
\de_{\rm Diff}\, \lvb \mu \nu \simeq \partial_\mu \ep_\nu \, , \quad\quad
\de_{\rm Diff}\,  \lsc \la \mu \nu \simeq 0 \, , \quad\quad\,
\label{linDiff}
\eeq
\beq
\de_{\rm LT}\, \lvb \mu \nu \simeq \partial_\mu \ep_\nu \, , 
\quad\quad
\de_{\rm LT}\, \lsc \la \mu \nu \simeq 0 \, ,
\quad\quad
\label{linLT}
\eeq
\beq
\de_{\rm LLT}\, \lvb \mu \nu \simeq \ep_{\mu\nu} \, , \quad\quad
\de_{\rm LLT}\, \lsc \la \mu \nu \simeq \partial_\la \ep_{\mu\nu} \, ,
\label{linLLT}
\eeq
where the Diffs use $\xi^\mu = \ep^\mu$.
When excitations of this form are inserted into~\rf{linR} and~\rf{linT},
$R_{\ka\la\mu\nu}$ and $T_{\la\mu\nu}$ both vanish at first order.  
Since the gauge modes have the form of the ten excitations
$\ep_\mu$ and $\ep_{\mu\nu} $,
the fact that they disappear at the linearized level means
that the consistency conditions stemming from~\rf{JYnogo}
generally cannot be satisfied unless constraints are imposed 
on the torsion and curvature or on the backgrounds themselves, 
either of which renders the framework useless.  

This argument extends as well to higher-dimensional operators
in the pure-gravity action,
which are coupled to background fields.
In a first-order treatment such terms consist of operators 
with one or more covariant derivatives
acting on the curvature or torsion.
However, when the action is linearized,
the covariant derivatives reduce
to partial derivatives at first order, for example,
\beq
D_\al D_\be R_{\ka\la\mu\nu} \simeq \partial_\al \partial_\be R_{\ka\la\mu\nu} \, ,
\label{Dtopartial}
\eeq
and likewise with the linearized torsion.
As a result, all the potential gauge degrees of freedom
drop out of these higher-dimensional operators as well.

The end result is that a perturbative pure-gravity approach with explicit breaking
in EC theory is generally not useful or is inconsistent.
See~\cite{akzl21a} as well for additional arguments and examples 
that reach the same conclusion.

Despite the breakdown of theoretical consistency
at the perturbative level with explicit breaking,
experiments can still conduct tests of gravity using the 
version of the SME that includes nondynamical backgrounds.
Any detection of a signal due to explicit breaking in a model
that does not evade the no-go results would then have to be 
interpreted as giving evidence of a geometry
that goes beyond Riemann or Riemann-Cartan geometry,
such as Finsler geometry.
This is the approach taken in~\cite{akzl21a},
while~\cite{akzl21b} examines a number of experimental tests 
involving both the pure-gravity and matter-gravity sectors.

The pure-gravity sector also includes interactions between
gravity waves and fixed background fields.
In general, the effects of spacetime symmetry breaking on
gravity waves can be investigated using a variety of approaches.
See, for example,~\cite{ssw80,fgnpps07,myw12,akmm16,akmm18,mm2019,xgs221,aa1572022,eisfv23}.
However, in the context of effective field theory, as is being considered here,
a perturbative approach using linearized gravity coupled to backgrounds 
of the form $\bar k_Y$ can be used~\cite{akmm16,akmm18,mm2019}.
In this framework, completely generalized operators consisting of
partial derivatives acting on metric excitations $h_{\mu\nu}$
in a Minkowski background coupled directly to $\bar k_Y$ are included,
as opposed to restricting to couplings with only the 
linearized curvature tensor.
To avoid conflicts with translation invariance,
it is assumed that the backgrounds $\bar k_Y$ are
constant or approximately constant.
At the linearized level,
breaking of Diffs becomes breaking of a gauge symmetry,
while breaking of Lorentz symmetry becomes global
breaking in the Minkowski background.
In this context, the linearized metric includes
additional degrees of freedom associated with 
gauge breaking under Diffs.
However, in general,
they may not appear in such a way that evades the no-go results.
Thus, any detection of spacetime breaking from gravity waves,
which breaks Diffs but does not evade the no-go conditions,
would be indicative as well
of a geometry that goes beyond Riemann or Riemann-Cartan.

For additional applications with explicit breaking, see,
for example~\cite{obn21,rs21,rss22,nan22}.

\subsection{Matter-Gravity Sector}

Setting aside the potential and pure-gravity terms,
the action term for the matter-gravity sector is $S_{g,m , \bar k_Z} $,
which contains the fixed backgrounds $\bar k_Z$.
The dynamical equations of motion for the vierbein and spin connection are 
\bea
G^{\mu\nu} = \ka \bar T_e^{\mu\nu} \, ,  \quad\quad\quad
\label{EinsteinEqkbarM} \\
\hat T^{\la\mu\nu}  
=   - \ka \bar S_{\om \, \pt{\mu} }^{\pt{\om \, } \la\mu\nu} \, ,
\quad
\label{TorsionEqkbarM} 
\eea
where for simplicity the matter fields are put on shell.
Comparing these to Eqs.\ \rf{EinsteinEq} and~\rf{TorsionEq}
shows that they have the same form as in EC theory except that the energy-momentum 
and spin density tensors now have bars over them, indicating that they depend 
on the background fields, $ \bar k_Z$.

The energy-momentum and spin density for ordinary matter are contained
in $\bar T_e^{\mu\nu}$ and $\bar S_{\om \, \pt{\mu} }^{\pt{\om \, } \la\mu\nu}$.
The background $\bar k_Z$ contributes to $ \bar T_e^{\mu\nu}$ as well
when it couples to both matter fields and the vierbein.
Similarly, $\bar k_Z$ contributes to 
$\bar S_{\om \, \pt{\mu} }^{\pt{\om \, } \la\mu\nu}$
when it couples to both matter fields and the spin connection.
The latter contributions arise, for example,
when covariant derivatives act on the matter fields,
such as $D_\mu \ps$ or $D_\mu A_\nu$,
and then also couple with $\bar k_Z$,
Thus all terms with couplings to $\bar k_Z$ in the matter-gravity sector
combine with both gravitational and matter fields,
which implies that in regions of spacetime where matter is absent, 
$ \bar T_e^{\mu\nu}$ and $\bar S_{\om \, \pt{\mu} }^{\pt{\om \, } \la\mu\nu}$ both vanish.
From~\rf{TorsionEqkbarM}, it follows that the torsion vanishes
as well in the absence of matter,
and therefore in vacuum the theory is no different from EC theory.

Assuming the no-go results are evaded, the Noether identities for
observer LTs and LLTs in this case are given in~\rf{LTNoether2obsTe} 
and~\rf{LLTNoether2obsTe},
which are consistent with the Bianchi identities
and the equations of motion in~\rf{EinsteinEqkbarM}
and~\rf{TorsionEqkbarM}.
Note how the identities 
in~\rf{LTNoether2obsTe} and~\rf{LLTNoether2obsTe} 
match the identities~\rf{LTNoether2} and~\rf{LLTNoether2}
in EC theory when bars are placed over $ \bar T_e^{\mu\nu}$ 
and $\bar S_{\om \, \pt{\mu} }^{\pt{\om \, } \la\mu\nu}$.

The condition in~\rf{nogo2} that must hold  to
evade the no-go results in this case is
\beq
\int d^4x \, e \,  J^Z \, \de \bar k_Z = 0 \, .
\label{nogoJZ}
\eeq
where $\de \bar k_Z$ can be observer Diffs, LTs, or LLTs 
of the backgrounds $\bar k_Z$.

\subsubsection{Energy-Momentum}

Because of the similarity with EC theory,
and assuming the no-go results can be evaded,
the same procedure as described in
Section~\rf{EMeffconsv} can be applied here as well.
An effective energy-momentum tensor with Belinfante-Rosenfeld form
can be defined as in \rf{Teff},
but with bars added to the energy-momentum and spin density so that
\beq
\bar T_{\rm eff}^{\mu\nu} = \bar T_e^{\mu\nu} 
- \half \tilde D_\si (\bar S_{\om  }^{\pt{\om}\si\mu\nu} + \bar S_{\om  }^{\pt{\om}\mu\nu\si} + \bar S_{\om  }^{\pt{\om}\nu\mu\si} ) \, .
\label{Teffbar}
\eeq
The Einstein tensor can again be divided into a Riemann part and
a non-Riemann part,
where on shell the torsion in the non-Riemann part can be written in terms of 
$\bar S_{\om \, \pt{\mu} }^{\pt{\om \, } \la\mu\nu}$.
In the limit where quadratic contributions 
$\sim \ka^2 \bar S_{\om  }^2$ can be neglected,
the effective energy-momentum tensor is 
\beq
\bar T_{\rm eff}^{\mu\nu} \simeq \bar T_e^{(\mu\nu)} 
- \half \tilde D_\si (\bar S_{\om  }^{\pt{\om}\mu\nu\si} + \bar S_{\om  }^{\pt{\om}\nu\mu\si} ) \, ,
\quad\quad\,\,\,\,
\label{Teffbar2}
\eeq
and the effective Einstein equation reduces to 
$\tilde G^{\mu\nu} \simeq \ka \bar T_{\rm eff}^{\mu\nu}$.
From this it follows that as long as the no-go results are evaded
$\tilde D_\mu \bar T_{\rm eff}^{\mu\nu} \simeq 0$
and $T_{\rm eff}^{\mu\nu} = T_{\rm eff}^{\nu\mu}$
must hold at leading order in a perturbative treatment.
This results in a perturbative theory that is effectively Riemann.

In particular, in a flat spacetime limit with negligible torsion,
$\partial_\mu T_{\rm eff}^{\mu\nu} \simeq 0$ would need to hold.
If a background $ \bar k_Z$ has spacetime dependence, 
this could lead to violations of global translation invariance 
and energy-momentum conservation.
A breakdown of energy-momentum conservation in a flat torsionless limit is
problematic as well, because no measurements detect such a violation.
For these reasons, it is assumed here that $ \bar k_Z$ is constant or
very close to constant over relevant distance scales,
so that $\partial_\mu \bar k_Z \simeq 0$ holds.
In this case, as long as the no-go results can be evaded,
the Noether identities and equations of motion are consistent,
and the theory closely parallels EC theory.

\subsubsection{No-go Results}

To evade the no-go results,
the identities that follow from~\rf{nogoJZ} must be solvable
when $\de \bar k_Z$ are given as observer Diffs, LTs, and LLTs.
In general, it is possible that these conditions can hold nonperturbatively,
since the current $J_{m}^Y$ is not typically gauge invariant.  
Thus, the extra degrees of freedom in the vierbein do not 
necessarily disappear and can take values that satisfy 
the consistency conditions in~\rf{nogoJZ}.
However, to fully evade the no-go results,
solutions must exist for the extra modes,
without placing additional constraints on the background fields.
Note that there are no additional degrees of freedom in the torsion,
which according to~\rf{TorsionEqkbarM} is fully determined
by the spin density.

Arguments concerning consistency at the level of perturbation theory
are less conclusive in the matter-gravity sector
compared to the pure-gravity sector.
Indeed, there are known examples of theories
with explicit breaking in the matter-gravity sector that evade the
no-go results.  See, for example,~\cite{rbhbyw19,rbyySym21,akzl21a}.
It is also the case that the currents $J^Y$ typically contain
both gravitational and matter fields coupled together,
which can make linearizing in a systematic way ambiguous.
For example, 
if small matter excitations only couple with the
zeroth-order vierbein $\vb \mu a \simeq \de_\mu^a$
and spin connection $\nsc \mu a b \simeq 0$,
then the extra degrees of freedom are generally suppressed.  
At the same time,
even with quadratic couplings retained in a perturbative treatment,
the general arguments given in~\cite{akzl21a} show that the
consistency conditions stemming from~\rf{nogoJZ}
can run into experimental constraints that conflict with a particular model.
For example, the experimental sensitivities that are relevant  
for $J^Z$, $\bar k^Z$, vierbein excitations, and matter fields
might be orders of magnitude apart and therefore incompatible.
This can result in a theory being perturbatively inconsistent
and not useful to use as a phenomenological framework.

\subsubsection{Constant Vector Background in Matter-Gravity Sector}

As concrete examples,
consider theories where the background $\bar k_Z$
is either a constant spacetime vector $\bar b_\mu$
or a constant local vector $\bar b_a$.
To first order in a perturbative approach,
with $\vb \mu a = \de_\mu^a$ and $\nsc \mu a b = 0$ at zeroth order,
the observer transformations under Diffs, LTs, and LLTs,
with infinitesimal parameters $\xi^\mu = \ep^\mu = \de^\mu_a \ep^a$ 
and $ \ep_{a}^{\pt{a} b}$, are given in this case as:
\beq
\de_{\rm Diff}\, \bar b_\mu \simeq (\partial_\mu \ep^\nu) \bar b_\nu \, , \quad\quad
\de_{\rm LT}\, \bar b_\mu \simeq (\partial_\mu \ep^\nu) \bar b_\nu \, , \quad\quad
\de_{\rm LLT}\,  \bar b_\mu \simeq 0 \, ,
\quad\quad
\label{linbmu}
\eeq
\beq
\de_{\rm Diff}\, \bar b_a \simeq 0 \, , \quad\quad
\de_{\rm LT}\, \bar b_a \simeq 0 \, , \quad\quad\quad
\de_{\rm LLT}\,\bar b_a \simeq - \ep_{a}^{\pt{a} b} \bar b_b \, .
\label{linba}
\eeq
Here, the constant spacetime components $\bar b_\mu$, 
obeying  $\partial_\nu \bar b_\mu \simeq 0$,
break particle Diffs and LTs but not particle LLTs
when they couple to dynamical quantities.
This is consistent with the relation in~\rf{transfrel},
which shows that Diffs and LTs are equal when LLTs vanish.
At the same time,
the constant local background $\bar b_a$, with $\partial_\nu \bar b_a = 0$,
breaks particle LLTs, 
but at leading order it does not break particle Diffs or LTs
in contrast to the result in~\rf{transfrel}.
Notice, however, that at second order in small quantities,
$\de_{\rm LT} \, \bar b_a = \ep^c \ivb \mu c D^{(\om)}_{\, \mu} \bar b_a
= \ep^c \ivb \mu c \nsc \mu a b  \bar b_a$,
as in~\rf{batrans}.
Hence, if $\nsc \mu a b \ne 0$ at first order,
the relation in~\rf{transfrel} would be applicable,
and it would confirm that LLTs and LTs are directly linked when Diffs vanish.

As this example illustrates, a theory with background $\bar b_\mu$ 
violates different symmetries than a theory with $\bar b_a$.
It also shows that a constant background can break particle LTs
without necessarily breaking global translations in a flat torsionless limit,
since $\partial_\nu \bar b_\mu = 0$ and $\partial_\nu \bar b_a = 0$.
The conditions that must hold to evade the
no-go results are the identities derived from~\rf{nogoJZ}, 
with  $\de \bar b_\mu$ or $\de \bar b_a$ given as 
infinitesimal Diffs, LTs, and LLTs (as defined in~\rf{linbmu} and~\rf{linba}) must hold.

For the case of constant $\bar b_\mu$, 
LLTs are unbroken to leading order,
while Diffs and LTs have the same form of breaking with $\xi^\mu = \ep^\mu$.
The resulting identity under LTs is 
\beq
[(D_\mu - T^\la_{\pt{\la} \la\mu}) J^\mu] \bar b_\nu = 0 \, .
\label{JbumucondZ}
\eeq
Thus, unless $(D_\mu - T^\la_{\pt{\la} \la\mu}) J^\mu = 0$,
the theory is inconsistent and the no-go results hold.
In a limit with weak gravity and negligible torsion,
the condition in~\rf{JbumucondZ} reduces to $\partial_\mu J_Y^\mu \simeq 0$.
In models where the current is approximately conserved in this manner,
consistency can hold.
However, in general ,
consistency would require cancelations between matter fields,
gravitational fields, and the background $\bar b_\mu$ to occur,
which generally have very different experimental sensitivities
as argued in~\cite{akzl21a}.

For constant $\bar b_a$,
with broken LLTs,
the conditions that must hold are
\beq
J^a \bar b^b - J^b \bar b^a = 0 \, .
\label{LLTJYcond}
\eeq
With LLTs, the extra gauge degrees of freedom are the antisymmetric
components in the vierbein,
and these are not sufficient at leading order in $J^a_Y$ to make~\rf{LLTJYcond} 
hold for generic values of a constant vector $\bar b_a$.
Thus, a no-go result holds at the perturbative level.  

In summary, in theories with constant or nearly constant explicit-breaking
backgrounds $\bar b_\mu$ or $\bar b_a$ in interaction with a matter-gravity current,
the question of whether no-go results can be evaded depends
in a case-by-case manner on the type of current and level of perturbation
theory that is used.
However, at leading order in perturbation theory,
the result for most models is 
either that the no-go conditions hold or that experimental constraints 
imply that they cannot hold.

\subsubsection{St\"uckelberg Approach}

Since a St\"uckelberg approach is commonly used in gravity
theories with explicit breaking of Diffs in Riemann space~\cite{ags03},
it is examined here in Riemann-Cartan space.
For simplicity, the example of a constant background vector 
$\bar b_\mu$ is considered again.
The technique involves introducing a set of St\"uckelberg 
scalar fields $\ph^a$ and using them to replace $\bar b_\mu$ as
\beq
\bar b_\mu \rightarrow (\partial_\mu \ph^a) \bar b_a \, ,
\label{buStuck}
\eeq
where $\bar b_a$ is constant and where the scalars
$\ph^a$ are dynamical fields.
This procedure (often called a trick) restores the broken Diffs.
At the same time, since the scalars $\ph^a$ are dynamical,
there are equations of motion for them that must hold.

In the St\"uckelberg approach,
Diffs are spontaneously broken by the
vacuum value for $\ph^a$,
which is given as
$\vev{\ph^a} = \de_\nu^a x^\nu$.
When $\ph^a$ takes this value, 
$(\partial_\mu \ph^a) \bar b_a = \bar b_\mu$, 
reproducing the fixed background $\bar b_\mu$.
The idea then is that any fixed nondynamical field that explicitly breaks 
Diffs can be reproduced as a gauge-fixed vacuum solution in a
theory with spontaneous breaking caused by the St\"uckelberg fields.

Making the substitution~\rf{buStuck} in the 
matter-gravity action term
changes it to a new action in terms of $\ph^a$:
\bea
S_{g,m , \bar k_Z}  = \int d^4x \, e \, J^\mu \bar b_\mu 
\quad\quad\quad\,\,\,
\nonumber \\
\rightarrow \int d^4x \, e \, J^\mu (\partial_\mu \ph^a) \bar b_a \, .
\label{Sstuck}
\eea
Varying $\ph^a$,
using integration by parts, and discarding a boundary term,
gives the result
\beq
[(D_\mu - T^\la_{\pt{\la} \la\mu}) J^\mu] \bar b_a = 0 \, .
\label{JbumucondYStuc}
\eeq
Therefore, the condition $(D_\mu - T^\la_{\pt{\la} \la\mu}) J^\mu = 0$ holds
as a result of the equations of motion for $\ph^a$, 
illustrating that the St\"uckelberg approach still works when there is torsion.

However, regardless of whether the condition $(D_\mu - T^\la_{\pt{\la} \la\mu}) J^\mu = 0$
holds as the result of invariance under observer Diffs or LTs or as the result
of using a St\"uckelberg approach,
the consistency of a theory still depends in both cases on whether this condition
can hold for generic constant vectors $\bar b_\mu$.
If not, the no-go result applies, and the theory is inconsistent.
Thus, while the St\"uckelberg method can be useful in exploring theories with 
explicit-breaking background fields for which the no-go results are evaded,
it does not provide a way to make a theory that is inconsistent into
one that is consistent.

%%%%%%%%%%%%%%%%%%%%%%%%%%%%%%%%%%%%%%%%%%%%%%%%%%%%%%

\section{Spontaneous Breaking} \label{sec5}

Spontaneous breaking of spacetime symmetry has been widely investigated both theoretically and 
experimentally~\cite{sme1,sme2,sme3,akgrav04,rbsme,JT14,RB14,hees2016,aknr-tables,DM2005,Will2002}.
Specific mechanisms for how spontaneous breaking can occur have 
been identified and explored~\cite{akss1,akss2,akss3},
and examples and properties of the NG modes that can arise have been 
studied~\cite{jdb63,yn68,prp66,hco69,kt02,scel2004,bmg2004,bb1,bb2,ee07,akrp05,cfjn08,akrp09,
smchtikw09,abk10,jalfu10,zbovk13,caelfu15,cah16,caerp20}.
In the SME, both the pure-gravity and matter-gravity sectors have been
constructed in EC theory~\cite{akgrav04}.
In addition, perturbative frameworks in Riemann space with spontaneous breaking
have been developed.
In the pure-gravity sector,
these include a post-Newtonian framework~\cite{qbak06}  and a
linearized perturbative framework suitable for gravity waves~\cite{akmm16}.
In the matter-gravity sector,
techniques leading to the construction of a consistent 
perturbative framework have been found~\cite{akjt1,akjt2}.   
These different frameworks have been used to analyze a wide
range of experimental tests of spacetime symmetry in gravity theories~\cite{aknr-tables}.
In addition, specific models exhibiting spontaneous spacetime symmetry have
been constructed and investigated~\cite{akss2,akss3,bb1,bb2}, 
including bumblebee models, where numerous applications have been explored
(for examples, see~\cite{objp05,ms10,ch14,ccps18,amnp19,zo20,caml21,rmjn21,dnopp21,fmbvmm21,lxls22,mxls23,klm23,lxms23,xls23,anpp23}).

With spontaneous breaking, the generic form of the action is given in~\rf{kECS},
and the equations of motion are given in~\rf{EinsteinEqkbar}--\rf{feqkbar}.
In general, the energy-momentum tensor does not have to be symmetric
or covariantly conserved when the torsion is nonzero.  
These equations show that the torsion can be sourced by
spin density from ordinary matter as well as by contributions that depends
on the backgrounds.

As originally shown in~\cite{akgrav04},
in an EC theory with spontaneous breaking of Diffs and LLTs,
the no-go results are evaded.
The same is true using LTs and LLTs as the basic symmetries,
since these yield an equivalent set of Noether identities.
The no-go results are evaded because the background fields, 
$\bar k_Y$, $\bar k_Y$, and $\bar k_Z$ are vacuum expectation values
of dynamical fields $K_Y$, $K_Y$, and $K_Z$,
which obey the equations of motion:
\beq
\fr {\de S_{g,\bar k_X}} {\de K_X} = 0 \, ,
\quad\quad\quad \fr {\de S_{g,m, \bar k_Z}} {\de K_Z} = 0 \, ,
\quad\quad\quad \fr {\de S_{g,m, \bar k_Z}} {\de K_Z} = 0 \, .
\label{kbareq}
\eeq
Thus, the backgrounds $\bar k_X$, $\bar k_Y$ , and $\bar k_Z$ 
are the vacuum solutions to these equations.
As a result, 
all three conditions in~\rf{nogo1} hold for the vacuum solutions.
Since each of the action terms are individually scalars under
observer LTs and LLTs,
the Noether identities for LTs in~\rf{LTNoether2obsU}--\rf{LTNoether2obsTe}
and for LLTs in~\rf{LLTNoether2obsU}--\rf{LLTNoether2obsTe} all hold.
These identities are consistent with each other, with the contracted forms
of the Bianchi identities in~\rf{ContractedBianchi1} and~\rf{ContractedBianchi2},
and with the equations of motion.

Notice that with spontaneous breaking,
there is nothing that prevents the breaking from happening in
more than one particle sector.
Thus, in principle $\bar k_X$, $\bar k_Y$ , and $\bar k_Z$ 
can all have nonzero values at the same time.
However, to avoid potential issues with breakdown of global translation invariance and
energy-momentum conservation in a flat limit,
the backgrounds $\bar k_X$, $\bar k_Y$, and $\bar k_Z$ can be approximated 
as constant or nearly constant on relevant experimental distance scales. 

When excitations about the vacuum are included,
the symmetry becomes hidden,
but NG modes and massive Higgs-like excitations combine
with the backgrounds and other dynamical excitations
to keep the symmetry unbroken in the action,
and the Eqs.\ in~\rf{kbareq} continue to hold.
The counting of degrees of freedom in the vierbein
and spin connection is unchanged from the case of EC theory,
and all ten of the spacetime symmetries, consisting of LLTs and either Diffs or LTs,
can be used to gauge away ten degrees of freedom.
The torsion does not propagate,
and on shell it is fixed by the spin density.

Just as in EC theory,
an effective theory can be found as described in Section~\rf{EMeffconsv},
where the curvature is split into Riemann and non-Riemann parts,
and the torsion is eliminated on shell using the equations in~\rf{TorsionEqkbar}.
The relevant equations in the effective theory are found by making the replacements
\beq
\ka T_e^{\pt{e}\mu\nu} \rightarrow 
( - \bar {\cal U}^{\mu\nu} -  \bar {\cal G}^{\mu\nu} + \ka \bar T_e^{\pt{e}\mu\nu} ) \, ,
\label{sub1}
\eeq
\beq
\ka S_{\om \, \mu\al\be} \rightarrow ( \bar {\cal T}_{\mu\al\be} +  \ka \bar S_{\om \, \mu\al\be} ) \, ,
\quad\quad\quad
\label{sub2}
\eeq
in Eqs.~\rf{Rcurvetilde} through~\rf{GtildeSw2}.
In a limit where the torsion and background fields are weak,
so that quadratic terms in $( \bar {\cal T}_{\mu\al\be} +  \ka \bar S_{\om \, \mu\al\be} )$
can be neglected,
the curvature becomes Riemann,
and the Einstein equations reduce to
$\tilde G^{\mu\nu} \simeq \ka \bar T_{\rm eff}^{\mu\nu}$.
It follows from this that $\tilde D_\mu \bar T_{\rm eff}^{\mu\nu} \simeq 0$ 
and $\bar T_{\rm eff}^{\mu\nu} \simeq \bar T_{\rm eff}^{\mu\nu}$ hold
in a weak-torsion limit.
As the bar indicates, $\bar T_{\rm eff}^{\mu\nu}$ in general depends on
the backgrounds $\bar k_X$, $\bar k_Y$ , and $\bar k_Z$ .
Nonetheless, with spontaneous breaking,
the energy-momentum is covariantly conserved in
an effective theory with Riemann curvature in a weak-torsion limit,
just as it is in EC theory with no symmetry breaking.

\subsection{Bumblebee Models}

Bumblebee models are useful for studying how spontaneous breaking of
spacetime symmetry can occur~~\cite{akss1,akss2,akss3,akgrav04,bb1,bb2}.
It allows various features and properties of the symmetry breaking to be explored,
including how NG modes and Higgs-like modes can appear, 
and whether a Higgs mechanism might occur.

The bumblebee field has spacetime components $B_\mu$ that are
connected to local components $B_a$ by the vierbein,
so that $B_\mu = \vb \mu a B_a$.
Its defining feature is that it has a potential $V$ that has a minimum when 
$B_\mu$ and the vierbein have nonzero vacuum values, 
which spontaneously break Diffs and LLTs.
In addition to having an Einstein Hilbert term,
the action for a bumblebee model can take a range of forms,
with different kinetic terms and with either minimal or nonminimal couplings.

The simplest form to consider is where there are only minimal couplings and where
the kinetic term has a Maxwell form,
in which case the bumblebee Lagrangian can be written as
\beq
{\cal L}_B = - \fr 1 4 B_{\mu\nu} B^{\mu\nu} - V(B_\mu B^\mu + b^2) \, .
\label{LBB}
\eeq
Here, $B_{\mu\nu}$ is the field strength,
which in Riemann space can be defined using covariant derivatives,
which simply reduce to partial derivatives.
However, in EC theory the torsion enters when covariant derivatives are used:
\beq
B_{\mu\nu} = D_\mu B_\nu - D_\nu B_\nu
= \partial_\mu B_\nu - \partial_\nu B_\mu - T^\la_{\pt{\la}\mu\nu} B_\la \, .
\label{Bmunu}
\eeq
For an electromagnetic field, including the term with torsion would
break local $U(1)$ gauge symmetry, and therefore
a definition in terms of only partial derivatives would be preferable.
However, in bumblebee models, the potential $V$ breaks local $U(1)$ invariance,
and therefore either definition of $B_{\mu\nu}$ can be considered.
In the case where covariant derivatives are used, and the torsion is included,
contributions to the spin density can occur.

The potential $V$ is given as a function of the combination $(B_\mu B^\mu + b^2)$,
where $b$ is a constant.
One possibility is to define $V$ as a smooth quadratic function,
\beq
V(B_\mu B^\mu + b^2) = \fr 1 2 \la (B_\mu B^\mu + b^2)^2 \, ,
\label{VBB}
\eeq
where $\la$ is a constant.
This potential has a minimum when
\beq
V^\prime = \la (B_\mu B^\mu + b^2) = 0 \, .
\label{VprimeBB}
\eeq
Thus, the components $B_\mu $ and $B_a$ as well as the vierbein
all have nonzero vacuum values when $V$ is at its minimum,
\beq
\vev{B_\mu} = \bar b_\mu \, , \quad\quad \vev{\vb \mu a} = \de_\mu^a \, .
\quad\quad  \vev{B_a} = \bar b_a \, ,
\label{BBvevs}
\eeq
Here, $\bar b_\mu$ and $\bar b_a$ are both assumed to be constant,
and a Minkowski background is assumed for the vierbein.
The vacuum metric is then given as $\vev{g_{\mu\nu}} = \et_{\mu\nu}$,
while the local metric is $\et_{ab}$.
The vacuum values for the background vector are related by 
$\bar b_\mu = \vev{\vb \mu a}  \bar b_a$,
which must obey $\bar b^\mu \bar b_\mu = \bar b^a \bar b_a = -b^2$
so that $V^\prime = 0$ holds.
Each of these vacuum values is fixed under particle spacetime transformations.
Thus, when one of them transforms under observer transformations, 
it spontaneously breaks the symmetry.

\subsubsection{Vacuum solution with $\vev{\nsc \mu a b} = 0$}

In EC theory, 
the vacuum solution for the spin connection can be chosen as
\beq
\vev{\nsc \mu a b} = 0 \, .
\label{vevBBsc}
\eeq
With this and the vierbein vacuum, $\vev{\vb \mu a} = \de_\mu^a$,
the vacuum values for the bumblebee field strength, spin density, torsion, 
energy-momentum, and curvature all vanish:
\beq
\vev{B_{\mu\nu}} = 0 \, ,
\quad\quad
\vev{ \bar S_{\om \, \pt{\mu} }^{\pt{\om \, } \la\mu\nu} } = 0 \, ,
\quad\quad
\vev{ T^\la_{\pt{\la}\mu\nu}} = 0 \, ,
\quad\quad
\vev{ \bar T_e^{\mu\nu}} = 0 \, ,
\quad\quad
\vev{ R^\ka_{\pt{\ka}\la\mu\nu}} = 0 \, .
\quad\quad
\label{vacvalues1}
\eeq
Together the vacuum values, $\bar b_\mu$, $\bar b_a$, and $\vev{\vb \mu a}$
spontaneously break all three spacetime symmetries: Diffs, LTs, and LLTs.
However, each individually spontaneously breaks different symmetries.

While the vierbein vacuum value spontaneously breaks all three symmetries,
the constant local background $\bar b_a$ spontaneously breaks three LLTs, 
but not Diffs or LTs.
The background $\bar b_a$ does not break LTs, 
because $D^{(\om)}_\mu \bar b_a =0$ to lowest order when $\bar b_a$ is constant 
and $\vev{\nsc \mu a b} = 0$,
and Diffs are not broken when $\bar b_a$ is constant.
In contrast, the spacetime vector $\bar b_\mu$ breaks Diffs and LTs, but not LLTs.
It breaks breaks LTs because the vierbein breaks LTs, and similarly for Diffs.
Notice, however, that a constant vector $\bar b_\mu$ only breaks one Diff,
where $\xi^\mu$ is in the same direction as the vacuum value $\bar b^\mu$.
Similarly, at leading order, $\bar b_\mu$ only breaks one LT where
$\ep^a$ is in the same direction as $\vev{\vb \mu a} \bar b^\mu$.
Finally, since $\bar b_a$ and $\vev{\vb \mu a}$ transform inversely
under observer LLTs, $\bar b_\mu$ does not transform under LLTs.

The fate of the NG modes in this bumblebee model was
examined in~\cite{bb1,bb2} for spontaneous breaking of Diffs and LLTs,
where it was found that massless NG modes for the broken LLTs can 
appear and propagate,
while the NG modes for Diffs disappear and do not propagate.
However, it is also possible to consider symmetry breaking where LTs and LLTs
are the independent transformations,
and to look at whether the formation of the NG modes 
is changed in any way from the case where Diffs and LLTs are independent.

As described in~\cite{bb1,bb2}, the excitations around the vacuum value in $B_\mu$
consist of a combination of excitations of both the local vector $B_a$ and the vierbein $\vb \mu a$,
which take the form
\beq
\vev{B_\mu} + \de B_\mu = (\vev{\vb \mu a} + \de \vb \mu a)(\vev{B_a} + \de B_a) \, ,
\label{BdeB}
\eeq
where $\de B_\mu$, $\de \vb \mu a$, and $\de B_a$ are the relevant excitations,
which are assumed to be small.
Thus, to first order,
\beq
\de B_\mu \simeq  ( \de \vb \mu a) \bar b_a + \vev{\vb \mu a} \de B_a \, .
\label{deB}
\eeq
To identify the physical degrees of freedom in the case where the NG modes are
associated with Diffs and LLTs,
ten gauge degrees of freedom must first be fixed.
This is necessary with spontaneous breaking because the symmetries
still hold when all the excitations are included.
One choice for the gauge fixing is to set the anti-symmetric components in
$\de \vb \mu a$ to zero, which fixes the six LLTs,
and then set $( \de \vb \mu a) \bar b_a = 0$,
which fixes the four gauge degrees of freedom under Diffs.
The result is that 
\beq
\de B_\mu \simeq \vev{\vb \mu a} \de B_a \, ,
\label{fixeddeB}
\eeq
where there are now four degrees of freedom remaining in $\de B_a$.
Since three LLTs are spontaneously broken by $\bar b_a$,
there are three NG modes that can appear.
These have the form of excitations generated by the broken LLTs,
which stay in the minimum of the potential $V$.
The remaining fourth mode is a massive mode,
which does not stay in the minimum of the potential $V$.
The three NG modes for LLTs
have the form $\de B_a \simeq - \ep_a^{\pt{a}b} \bar b_b$,
where the broken generators, $\ep_a^{\pt{a}b}$, become the fields for the NG excitations.
These obey $(\de B_a) \bar b^a = 0$,
while the massive mode in $\de B_a$ is an excitation along the direction 
of $\bar b_a$.

In $\de B_\mu$, as given in~\rf{fixeddeB},
it is only the NG modes for LLTs that appear,
while the NG mode for the broken Diff disappears.
Under the broken Diff, the NG mode would appear as an excitation
$(\partial_\mu \xi^\nu) \vev{\vb \nu a} \bar b_a$,
where in this case $\xi^\mu$ for the broken transformation would become
the field for the NG mode.
The reason it disappears is because gauge fixing $( \de \vb \mu a) \bar b_a = 0$
imposes a condition on any excitations that might arise as NG modes in the vierbein:
\beq
(\partial_\mu \xi^\nu) \vev{\vb \nu a} \bar b_a - \ep^a_{\pt{a}b} \vev{\vb \mu b} \bar b_a = 0 \, .
\label{gaugeFixDiffLLT}
\eeq
This condition locks the NG mode for the broken Diff to the NG modes for 
the broken LLTs, and they cancel in the vierbein excitations.
The net result is that only the three NG modes for the broken LLTs appear in $\de B_\mu$.

A corresponding analysis can be carried out starting from~\rf{deB} and finding the NG modes 
when it is LTs and LLTs that are spontaneously broken.
In this case, the gauge freedoms under LTs and LLTs must be fixed,
which can be accomplished by again setting the anti-symmetric components in
$\de \vb \mu a$ to zero,
and then using the four LTs to set $( \de \vb \mu a) \bar b_a = 0$.
The result is again that $\de B_\mu \simeq \vev{\vb \mu a} \de B_a$
consists of three NG modes for spontaneous breaking of LLTs
and one massive mode.
The gauge fixing in this case locks excitations having the 
form of LTs and LLTs in $( \de \vb \mu a) \bar b_a$,
imposing the condition that 
$(\partial_\mu \ep^a)  \bar b_a - \ep^a_{\pt{a}b} \vev{\vb \mu b} \bar b_a=0$.
Here, the first excitation $(\partial_\mu \ep^a)  \bar b_a$ is what would be 
the NG mode for the broken LT;
however, it is locked to the NG modes for the broken LLTs by the gauge fixing,
which therefore causes them to cancel in the vierbein excitations.
The net result is again that only the three NG modes for the broken LLTs
appear in $\de B_\mu$.

Thus, regardless of whether Diffs and LLTs are used or LTs and LLTs are used,
the only NG modes that propagate as physical modes are the three NG modes 
stemming from spontaneous breaking of LLTs.
No NG modes for either Diffs or LTs appear in the theory.

\subsubsection{Vacuum solution with $\vev{\nsc \mu a b} \ne 0$}

An alternative vacuum structure can be considered for the case of a constant background $\bar b_a$, 
which spontaneously breaks both LTs and LLTs in the local frame.
It occurs when the vacuum vierbein has a Minkowski solution, $\vev{\vb \mu a} = \de_{\mu}^a$,
and the spin connection also has a nonzero vacuum value,
\beq
\vev{\nsc \mu a b} \ne 0 \, .
\label{nonzerovevBBsc}
\eeq
The condition that the bumblebee vacuum is in the minimum of the potential,
obeying $V^\prime = 0$,
is still satisfied when $\vev{B_a} = \bar b_a$ and $\vev{B_\mu} = \vev{\vb \mu a} \bar b_a$.
Spontaneous breaking of LTs occurs in this case, because under observer transformations,
$\de_{\rm LT} \bar b_a = \ep^c \ivb \mu c D^{(\om)}_\mu \bar b_a$,
and therefore the vacuum solution obeys
\beq
\de_{\rm LT} \bar b_a = \ep^c \vev{\ivb \mu c} \vev{\om_{\mu a}^{\pt{\mu a} b}} \bar b_b \ne 0 \, .
\label{LTbbaravac}
\eeq
In this case, the constant vector $\bar b_a$ spontaneously breaks three LTs,
where the broken generators $\ep^a$ are transverse to $\bar b^a$.
Thus, three NG modes for the broken LTs are expected.
At the same time, spontaneous breaking of LLTs occurs, since 
$\de_{\rm LLT} \bar b_a = -\ep_a^{\pt{a}b} \bar b_b \ne 0$
under observer LLTs.
Thus, there are three broken LLTs
and three NG modes for LLTs are expected.

Notice that the spin connection is not forced to have a nonzero vacuum value
so that $V^\prime = 0$ holds.
Instead, it acquires a vacuum value spontaneously at the same time that $\vb \mu a$ and $B_a$ 
take vacuum values making $V^\prime =0$.
However, as long as $\vev{\nsc \mu a b}$ is constant,
in addition to $\bar b_a$ and $\vev{\vb \mu a}$ being constant,
then the bumblebee field strength, spin density, torsion, energy-momentum, and curvature can all have constant
nonvanishing vacuum values:
\beq
\vev{B_{\mu\nu}} \ne 0 \, ,
\quad\quad
\vev{ \bar S_{\om \, \pt{\mu} }^{\pt{\om \, } \la\mu\nu} } \ne 0 \, ,
\quad\quad
\vev{ T^\la_{\pt{\la}\mu\nu}} \ne 0 \, ,
\quad\quad
\vev{ \bar T_e^{\mu\nu}} \ne 0 \, ,
\quad\quad
\vev{ R^\ka_{\pt{\ka}\la\mu\nu}} \ne 0 \, .
\quad\quad
\label{vacvalues2}
\eeq
Thus, the bumblebee vacuum Lagrangian in this
case is a constant,
whereas it was zero when the spin connection
had a vacuum solution $\vev{\nsc \mu a b} = 0$.

Since the spin connection has mass dimension equal to one,
its vacuum value can be written as $\vev{\nsc \mu a b} \sim \bar w$
(dropping indices), where $\bar w$ has units of mass.
Here, $ \bar w$ sets the energy scale for the spin connection vacuum solution.
Similarly, $\vev{B_a} \sim \vev{B_\mu} \sim b$,
where $b$ also has mass units.
The vacuum bumblebee field strength is nonzero because of the torsion
contribution in~\rf{Bmunu},
which gives
\beq
\vev{B_{\mu\nu}} = - \vev{T^\la_{\pt{\la}\mu\nu}} \vev{\vb \la a} \bar b_a \label{vacBmunu} \, ,
\eeq
where the torsion vacuum value is
\beq
\vev{T_{\la \mu \nu}} = \vev{\om_{\mu\la\nu}} - \vev{\om_{\nu\la\mu}} \, .
\label{Torvev}
\eeq
This shows that the torsion has mass dimension one and scales as $\vev{T_{\la \mu \nu}} \sim \bar \om$,
while the field strength has mass dimension two and scales as $\vev{B_{\mu\nu}} \sim \bar \om b$.
The vacuum spin density $\vev{ \bar S_{\om \, \pt{\mu} }^{\pt{\om \, } \la\mu\nu} }$ is found
by varying the Lagrangian term $- \fr 1 4 B_{\mu\nu} B^{\mu\nu}$ with respect to the spin connection
and substituting in the vacuum solutions.
It has mass dimension three and scales as 
$\vev{ \bar S_{\om \, \pt{\mu} }^{\pt{\om \, } \la\mu\nu} }  \sim \bar \om b^2 $.
Since the torsion equation for the vacuum is $\vev{\hat T^{\la\mu\nu}} =  - \ka \vev{\bar S_{\om \, \pt{\mu} }^{\pt{\om \, } \la\mu\nu}} $,
it shows that on shell the torsion scales as $\vev{T_{\la \mu \nu}} \sim \ka \bar \om b^2$,
where the dimensional coupling $\ka \sim M_{\rm Pl}^{-2}$ has mass dimension minus two, 
with $M_{\rm Pl}$ equal to the Planck mass.
The fact that the torsion scales as both $\sim \bar \om$ and  $\sim \ka \bar \om b^2$
on shell shows that $\ka b^2 \sim 1$ or that $b \sim M_{\rm Pl}$,
and therefore $\vev{T_{\la \mu \nu}} \sim \bar \om$ and 
$\ka \vev{ \bar S_{\om \, \pt{\mu} }^{\pt{\om \, } \la\mu\nu} } \sim \bar \om$.
Similarly, the energy-momentum is found by varying $- \fr 1 4 e B_{\mu\nu} B^{\mu\nu}$
with respect to the vierbein.
The result is that $ \bar T_e^{\mu\nu}$ is dimension four and scales as $\sim {\bar \om}^2 b^2$.
However, $\ka \bar T_e^{\mu\nu} \sim {\bar \om}^2$.
Finally, the vacuum curvature, which has mass dimension two, scales as $\vev{ R^\ka_{\pt{\ka}\la\mu\nu}} \sim {\bar \om}^2$.

Notice how these equations set the bumblebee mass scale $b$ to the Planck mass,
but the scale for the spin connection vacuum $\sim \bar \om$ is left undetermined.
However, since the vacuum torsion scales as $\sim \bar \om$ and the vacuum curvature scales as $\sim {\bar \om}^2$,
on experimental grounds the parameter $\bar \om$ must be very small.
This would be consistent with $\vev{\nsc \mu a b}$ arising spontaneously as a small vacuum value.
While the potential term scales as $V \sim b^4 \sim M_{\rm Pl}^4$,
it is zero in the vacuum solution.
In contrast, the other operator terms in the action all scale as $\sim {\bar \om}^2 M_{\rm Pl}^2$.
Matching the scale of these other nonvanishing vacuum operators 
on experimental grounds to a cosmological constant term that scales as
$\sim \La M_{\rm Pl}^2$ indicates that an appropriate scale for the vacuum spin connection would be $\bar \om \sim \sqrt{\La}$,
and hence $\bar \om$ must be extremely small.

To examine the NG modes for spontaneous breaking with $\vev{\nsc \mu a b} \ne 0$,
consider again the excitations in~\rf{deB}.
Ten gauge symmetries must be fixed.  
The six LLTs can be fixed by setting the anti-symmetric components of the vierbein to zero,
and the four LTs can be chosen so that the excitations $(\de \vb \mu a) \bar b_a  = 0 $.
This leaves $\de B_\mu \simeq \vev{\vb \mu a} \de B_a$,
where $\de B_a$ consists of three NG modes for the broken LTs 
and three NG modes for the broken LTTs all acting together,
which are given as
\beq
\de B_a = \ep^c \vev{\ivb \mu c} \vev{\om_{\mu a}^{\pt{\mu a} b}} \bar b_b
- \ep_a^{\pt{a}b} \bar b_b \, ,
\label{deBaLT}
\eeq
where $\ep^a$ and $\ep_a^{\pt{a}b}$ become
the fields for the NG modes.
These combined excitations obey $(\de B_a) \bar b^a = 0$
and therefore remain in the minimum of the potential $V$.
The massive mode is along the direction of $\bar b_b$,
and is assumed to be highly suppressed due to the mass scale $b \sim M_{\rm Pl}$.

With the gauge freedom fixed by setting $(\de \vb \mu a) \bar b_a  = 0$,
a condition is imposed on any excitations arising as NG modes for
LTs and LLTs in the vierbein:
\beq
(\de \vb \mu a) \bar b_a = (\partial_\mu \ep^a) \bar b_a +  \ep^b \vev{\om_{\mu \pt{a} b}^{\pt{\mu} a}}  \bar b_a
+ \ep^b \vev{T^a_{\pt{a} b \mu}}  \bar b_a
- \ep^a_{\pt{a}b} \vev{\vb \mu b} \bar b_a = 0 \, .
\label{gaugefix}
\eeq
Substituting for the torsion, rearranging, and making insertions of the vacuum vierbein,
this condition can be rewritten as:
\beq
\vev{\ivb \mu a}(\partial_\mu \ep^b) \bar b_b 
- [\ep^c \vev{\ivb \mu c} \vev{\om_{\mu a}^{\pt{\mu a} b}} \bar b_b
- \ep_a^{\pt{a}b} \bar b_b] = 0 \, .
\label{gaugefix2}
\eeq
This form shows that the excitation $\vev{\ivb \mu a}(\partial_\mu \ep^b) \bar b_b$,
which involves the LT with $ \ep^a$ along the direction of $\bar b^a$, 
is locked to the combination of NG modes for the three LTs and the three LLTs,
which appear together in $\de B_a$ in~\rf{deBaLT}.
Thus, the condition in~\rf{gaugefix2} prevents a fourth NG mode for the LTs,
with $\ep^a$ parallel to $\bar b^a$,
from appearing in $\de B_\mu $.

Notice also that the locked excitation $(\partial_\mu \ep^b) \bar b_b$
takes a form in the spacetime frame that is the same as a Diff NG mode:
$ (\partial_\mu \xi^\nu) \bar b_\nu$,
where $\bar b_\nu = \vev{\vb \nu b} \bar b_b$ and the Diff generator is $\xi^\nu = \vev{\ivb \nu a} \ep^a$.
In fact, if the vacuum value $\vev{\om_{\mu a}^{\pt{\mu a} b}}$ is set equal to zero,
so that LTs are no longer broken,
then the condition in~\rf{gaugefix2} reduces to the condition that locks
the Diff NG mode to the NG modes from LLTs just as in the previous example.

The net result of this example, however, is that with $\vev{\nsc \mu a b} \ne 0$,
there are three excitations in $\de B_\mu $,
which consist of a combination of three NG modes for LTs and three NG modes for LLTs.
The fourth mode is the bumblebee massive mode,
which scales as $\sim b \sim M_{\rm Pl}$,
and therefore its excitations are heavily suppressed and are unlikely to matter at lower energies.

The nature of the full solution depends on which components of the spin connection acquire vacuum values.
This would determine which components of the vacuum values in~\rf{vacvalues2} are nonvanishing,
and how they combine to give an overall constant vacuum Lagrangian.
With the gravitational, NG, and massive mode excitations included,
the physical viability of such a theory could be evaluated.
Additional terms involving nonminimal couplings or additional couplings
to the torsion could be considered as well.  
See, for example,~\cite{iw16}).
In addition, adding terms that allow the spin connection to propagate
would allow investigation of the possibility of a Higgs mechanism
for the spin connection.
However, exploring possibilities like these
goes beyond the scope of the present work,
which is focused on the nature of the spontaneous breaking of spacetime symmetries,
the resulting vacuum structure, and the appearance of NG modes.

%%%%%%%%%%%%%%%%%%%%%%%%%%%%%%%%%%%%%%%%%%%%%%%%%%%%%%

\section{Discussion and Conclusions} \label{sec6}

In this paper, the processes of explicit and spontaneous spacetime symmetry 
breaking in EC theory when fixed background fields are present have
been reviewed and reexamined with a focus being placed on the roles of torsion and LTs.

The original investigation of spacetime symmetry breaking in gravity~\cite{akgrav04},
published 20 years ago, found that with explicit breaking no-go 
results consisting of inconsistencies 
between the Bianchi identities and the equations of motion can arise,
but that with spontaneous breaking the no-go results are evaded.
The construction of the gravity sector of the SME based on the idea of
spontaneous breaking of spacetime symmetries has provided a
phenomenological framework that has been used in numerous gravity tests.
These include both pure-gravity and matter-gravity tests,
where extremely high sensitivities to possible symmetry breaking have been attained.
In addition, implications of spontaneous breaking of spacetime symmetries
have been explored, including looking at various scenarios and mechanisms for the
symmetry breaking as well as addressing questions concerning the appearance of
NG modes, Higgs-like modes, and the possibility of a gravitational Higgs mechanism.

However, over the past decade, gravity theories with explicit breaking
of spacetime symmetry breaking have been investigated in more detail,
and it was found in certain cases that it is possible to evade the no-go results.
Seeing how this occurs involves looking at the relationships between
the Bianchi identities, Noether identities stemming from observer independence,
energy-momentum conservation, and the equations of motion.
It was found that the no-go results can most readily be evaded in
theories where the extra degrees of freedom in the vierbein due to
explicit breaking can appear and have nonperturbative interactions with matter 
and gravitational fields.
It is also the case that highly specific forms of background fields are
able to evade the no-go results,
while generic forms typically do not.
In addition, in perturbative treatments,
which are widely used in gravity tests,
the extra modes often disappear or are highly restricted so that
useful solutions evading the no-go results do not exist.
This makes post-Newtonian approaches not useful
when the symmetry breaking is explicit.
Nonetheless, to investigate the possibility of explicit breaking experimentally,
a generalization of the SME with gravity has been developed, 
where the interpretation is that if any violations are discovered
in cases where the no-go results are not evaded,
they would give evidence of a geometry that goes beyond Riemann or 
Riemann-Cartan, such as Finsler geometry.

Much of the progress that has been made in understanding 
the differences between explicit and spontaneous breaking
has relied on investigations in which Diffs and LLTs are broken,
while the breaking of LTs has been looked at to a lesser extent.
In addition, the effects of torsion have often been ignored, 
including the question of how to interpret the fact that energy-momentum
is not covariantly conserved when torsion is present.

The primary goal of this paper has been to revisit EC theory with
background fields included and to look more closely at the
effects of torsion and breaking of LTs in theories with either spontaneous 
or explicit breaking.
One immediate effect involving  LTs in theories with fixed backgrounds
is the possibility of violation of global translation invariance in 
a flat spacetime limit and breaking energy-momentum conservation.
For this reason, the investigation here has been limited to the case where
the backgrounds are constant or nearly constant on relevant distance scales,
which is an assumption that is often made in the SME. 
The idea is that if no-go results arise for a constant background,
then having a background with spacetime dependence would be
even more problematic.
Interestingly, it is possible for LTs to be broken by 
backgrounds that are constant,
where presumably breaking of global translations is then not an issue.
However, even with restriction to constant backgrounds, it is
found that the no-go results still typically occur with explicit breaking.
Moreover, even if a theory is not totally ruled out,
in a perturbative treatment, the no-go results can still
render it useless as a phenomenological framework.

There are noteworthy effects involving torsion that can occur in EC theory when
background fields are present.
One is that background fields can act as a source of spin 
density and torsion even in a vacuum.
However, for this to work with explicit breaking the no-go results
must be evaded,
while with spontaneous breaking there are no no-go results.
As long as the spin connection has a nonvanishing coupling to
a background field, spin density and torsion can result in a vacuum solution.
Another effect of torsion is that energy-momentum is not
covariantly conserved when it is present.
However, in theories with background fields,
it is found that the effects of torsion can be accounted for
in ways that parallel what happens in EC theory.
With no symmetry breaking in EC theory,
there are weak limits where the curvature is Riemann
and the spin density effectively acts like additional contributions to the energy-momentum.
This same interpretation can hold as well when background fields are present.
However, with explicit breaking, it can only hold in highly restricted cases, 
and it cannot hold at all if the no-go results are not evaded.
In contrast, with spontaneous breaking, the no-go results are evaded,
and thus the same interpretation of energy-momentum conservation 
when torsion is included can hold as in EC theory.

The effects of torsion and spontaneous breaking of LTs were looked at 
specifically for the case of a simple bumblebee model with a Maxwell kinetic term.
To consider the effects of torsion, 
a field strength defined with covariant derivatives that includes coupling
to the torsion was chosen.
Two scenarios were examined,
one where the vacuum value of the spin connection vanishes,
and the other where it spontaneously takes a constant value.
In the latter case, spontaneous breaking of LTs occurs in the local frame,
whereas such breaking does occur when the vacuum spin connection is zero.
The spin density, curvature, and energy-momentum density can all
have constant vacuum values when the spin connection has one.
A comparison of the NG modes that can occur with the two types
of vacuum structures reveals that the NG modes for LLTs are 
present as propagating degrees of freedom in both cases.
However, in the case where the spin connection has a vacuum value,
the bumblebee modes are due to a combination of both 
breaking of LLTs and LTs.
At the same time, in both scenarios, the NG mode due to spontaneous
breaking of the Diff or the LT along the direction of the bumblebee
background vector does not propagate.

Overall, the results found here continue to show that explicit breaking
of spacetime symmetry is much more unnatural and problematic 
compared to spontaneous breaking.  
Effects of torsion or breaking of LTs do not alter this conclusion.
The processes involved in spontaneous breaking of spacetime symmetry
are far more elegant and compatible with EC theory 
than the corresponding processes that occur with explicit breaking.

%%%%%%%%%%%%%%%%%%%%%%%%%%%%%%%%%%%%%%%%%%

%%%%%%%%%%%%%%%%%%%%%%%%%%%%%%%%%%%%%%%%%%
\end{document}